\documentclass[twocolumn,prl,superscriptaddress,floatfix,showpacs,nofootinbib]{revtex4-1}
\usepackage[utf8]{inputenc}
\usepackage{amsmath}
\usepackage{amssymb}
\usepackage{graphicx}

\makeatletter
\usepackage{graphics}
\usepackage{epsfig}
\usepackage{color}

\usepackage[normalem]{ulem}
\newcommand{\be}{\begin{equation}} \newcommand{\ee}{\end{equation}}
\newcommand{\bea}{\begin{eqnarray}} \newcommand{\eea}{\end{eqnarray}}

\makeatother

\begin{document}

\title{Many universality classes in an interface model restricted to non-negative heights}

\author{Peter Grassberger} \affiliation{JSC, FZ J\"ulich, D-52425 J\"ulich, Germany} \affiliation{MPI for the Physics
of Complex Systems, D-01187 Dresden, Germany}

\author{Deepak Dhar} \affiliation{Indian Institute of Science Education and Research,  Dr. Homi Bhabah Road, Pune,  411 008, India } 
\author{P. K. Mohanty} \affiliation{Indian Institute of Science Education and Research - Kolkata, Mohanpur, Nadia - 741 246, India}

\date{\today}

\begin{abstract}

We present a simple one dimensional stochastic model with  three control parameters and a surprisingly rich zoo of phase transitions. At 
each (discrete) site $x$ and time $t$,  an integer $n(x,t)$  satisfies a linear interface equation 
with added random noise. Depending on the control parameters, this noise may or may not satisfy the  detailed balance 
condition,  so that the growing interfaces are in the Edwards-Wilkinson (EW) or in the Kardar-Parisi-Zhang (KPZ) universality class.  In addition, there is also a
constraint $n(x,t) \geq 0$. Points $x$ where $n>0$ on one side and $n=0$ on the other are called ``fronts". These fronts 
can be ``pushed" or ``pulled", depending on the control parameters. For pulled fronts, the lateral spreading 
is in the directed percolation (DP) universality class, while it is of a novel type for pushed fronts, with yet another 
novel behavior in between. In the DP case, the activity at each active site can in general be arbitrarily large, in 
contrast to previous realizations of DP. Finally, we find two different types of transitions when the interface detaches 
from the line $n=0$ (with $\langle n(x,t)\rangle \to$ const on one side, and $\to \infty$ on the other), again with new 
universality classes. We also discuss a mapping of this model to the avalanche propagation in a 
directed Oslo rice pile model in specially prepared backgrounds.
\end{abstract}
\maketitle

\section{Introduction}

Simple low dimensional stochastic models -- either in equilibrium or out of equilibrium -- have been at the core
of statistical physics for a very long time. One reason is that they can -- if they are sufficiently simple and 
``natural" -- be used for bewildering wide ranges of phenomena. We might just mention the Ising model that was 
conceived as a model for magnets, but has found applications in liquid-gas critical points \cite{Reichl}
and in social dynamics \cite{Castellano},
to mention just a few. Another one is random walks and modifications thereof, which are  relevant not only for 
Brownian motion but also for finance \cite{Bouchaud} and for polymer configurations \cite{deGennes}. 
As final examples we have ordinary and directed percolation (DP). 
Both are basic models for the spreading of epidemics \cite{epidemics}, although the former has also found
application in the sol-gel transition \cite{Stauffer}, while the latter  was -- under the name of 
Reggeon Field Theory -- once considered as a core model for ultra-relativistic hadron collisions
\cite{Grass-Sund,Sugar}.

This property of a model being applicable in a wide variety of contexts is particularly pronounced in one 
dimension, where particle, spin, and 
interface models can be mapped onto each other. But this mapping can exist also in higher dimensions, 
where it was observed by Paczuski and  Boettcher \cite{Paczuski} that ``sandpile" models \cite{btw} can be 
mapped exactly onto models for interface depinning. 

In the present paper we shall introduce and discuss a very simple one dimensional (1D) fully discrete model. We shall formulate
it as a stochastic evolution model of an interface with point-like pinning centers and a lower barrier below 
which the interface cannot go. This interface can get pinned at the barrier, and when it detaches, it can either 
grow laterally (i.e., in more and more regions the height above the barrier changes from zero to $>0$), or in 
height. These two types of transitions are associated with different universality classes. We will also show that
the model can also be interpreted as a model for the spreading of avalanches in specially prepared backgrounds 
in the directed version \cite{Oslo-Dhar} of the Oslo ricepile model \cite{Oslo}. 

The lateral spreading of detached regions can -- depending on the 
control parameters -- be in the universality class of DP. But, while activity in the 
active phase is binary in all previous realizations of DP (sites can either be dead or active), in the present 
model it can be active to varying degrees, and this degree of activity is a slow variable. While DP can be used 
as a model for infections where reproduction of infectants is so fast that a classification of individuals as 
healthy/infected is sufficient \cite{epidemics}, it is not suitable for some helminth infections (e.g. with 
{\it Ascaris lumbricoides} \cite{helminth,helminth1,helminth2}) where parasites cannot reproduce within the infected 
individual, and the number of parasites in an individual is a slowly varying relevant parameter. Although we 
shall not go into detail, it is clear that our model is thus more suitable as a model of helminth infections 
than standard DP.

A formal definition of the model is  given in the next section. In Sec.~III, we will discuss the sector 
where the interface is so far above the barrier that the latter is not felt. In that case the model shows 
-- depending on parameters -- scaling in the Kardar-Parisi-Zhang (KPZ) \cite{KPZ} and Edwards-Wilkinson (EW) 
\cite{EW} universality classes, respectively. Interfaces which stay close to the barrier, i.e. for which 
$\langle n(x,t)\rangle \to const < \infty$ for $t\to\infty$  (or,  in the tent phase, where  $\langle n(x,t)\rangle$ remains bounded by a time-independent constant  for most $x$ at any $t$)
are discussed in Sec.~IV, while the transition between interfaces attached to the barrier to interfaces which 
become detached, i.e. for which $\langle n(x,t)\rangle \to\infty$ for $t\to\infty$, is treated in 
Sec.~V. Finally, a mapping to propagation of avalanches in the directed Oslo rice pile model on specially 
prepared backgrounds is discussed in Sec.~VI.    Section VII summarizes our results.

\section{ Definition of the model }

We consider a 1-d lattice of size $2L$ with periodic boundary conditions and synchronous updating. At time 
$t=0$ a non-negative integer $n(x,t)\geq 0$ is attached to each even site $x$. Odd (even)  sites are updated
only for odd  (even) times, with  the evolution rule
\begin{equation}
    n(x,t+1) = \frac{1}{2} n_{in}(x,t+1) + \eta (n_{in}(x,t+1))  \label{eq:1d}
\end{equation}
where
\begin{equation}
    n_{in}(x,t+1)= n(x-1,t) +   n(x-1,t). 
\end{equation}
Here, $\eta (n_{in})$ is the simplest non-trivial noise term that depends only locally on $n$, preserves 
the non-negativity and integer nature of $n(x,t)$, and  leaves the zero state $n(x,t)=0$ invariant. 
This leaves us with
\begin{eqnarray}
    \eta(n) = \left\{ \begin{array}{r@{\quad}l@{\quad}l}
                       0      &  {\rm for}\;\;  n=0 & \\
                      \pm\frac12 & \mbox{with prob's.} \;(q,1-q) \;\mbox{for odd}\;  n  \\
                     \pm 1,0  & \mbox{with prob's.} \;(p_{\pm 1}, p_0)\;\mbox{for even}\; n>0  \\
                 \end{array} \right.    \label{noise}
\end{eqnarray}
where $p_{-1}+ p_0+ p_1 = 1$.

In the following we shall usually interpret $n(x,t)$ as the height of an interface without overhangs, 
although we shall also  mention some other 
interpretations. We shall consider only initial conditions where
$n(x,t)$ is either flat or has finite variance, i.e. we shall consider neither  interfaces which have 
infinite roughness already initially, nor circular interfaces as in the Eden model \cite{Barabasi}.

Before going on, we point out that the average instantaneous velocity of an interface 
${\cal I}_{t-1}=\{n(x,t-1)\}$,
\be
    v(a,t) = \frac{1}{L} [\sum_{x:x+t={\rm even}} \langle n(x,t) \rangle - \sum_{x:x+t={\rm odd}} n(x,t-1)],
    \label{naive}
\ee
can be calculated from $m_{\rm odd}(t)$ and $m_{\rm even}^+(t)$, which are the number of sites where 
$n_{in}(x,t)$ is odd or even and positive. Indeed it follows straightforwardly from Eqs.~(\ref{eq:1d}) and 
(\ref{noise}) that 
\be
   v(a,t) = \frac{1}{L}[(q-1/2)\langle m_{\rm odd}(t)\rangle  
	   + (p_1-p_{-1}) \langle m_{\rm even}^+(t)\rangle].      \label{m}
\ee
This is extremely useful in simulations, since Eq.~(\ref{m}) gives always smaller statistical
fluctuations then the `naive' use of Eq.~(\ref{naive}), in particular when $q$ is close to 1/2 
\cite{Grass-KPZ}. 
 The reason is that
relative fluctuations in $m_{odd}(t)$  and $m_{even}^+(t)$ are much less than in $\eta (x,t).$ In fact,
Eq.~(\ref{m}) is a variance-reduced estimator similar to those used in 
\cite{grassberger_percol_high-d,grassberger_log4d,Foster,JfWang}. Its use was crucial for obtaining 
the results of \cite{Grass-KPZ} and of the present paper.

\section {Interfaces where $n(x,t)$ is strictly positive for all $x$ and $t$}

If the interface is everywhere strictly above the barrier, the barrier at $n=0$ is ineffective,  and it 
remains so, if the mean velocity of the interface is positive, provided that statistical fluctuations 
don't push it down to $n=0$. Assume that the $p_i$'s are such that $v>0$ for $q=1$. Then the barrier
is ineffective for all times, if we started sufficiently high so that these fluctuations can be 
neglected. 

As we decrease $q$, keeping the $p$'s constant, the velocity can only decrease.  If it becomes negative,  the moving interface eventually comes in contact
with the barrier. We call this value of $q$ 
where $v=0$ the critical value $q_{c,a}(p_0,p_1)$, where ``$a$" stands for $a$bsence of barrier. If we 
decrease $q$ further, the interface will finally, i.e. for sufficiently small $q$, get stuck at $n=0$. 
But there might be an interval of $q$ between $q_{c,a}(p_0,p_1)$ and a second threshold 
$q_{c,b}(p_0,p_1)$ (where ``$b$" stands for $b$arrier) where it fluctuates above $n=0$, without getting 
completely absorbed or being able to detach completely. Indeed, $q_{c,b}$ might be either smaller or 
larger than $q_{c,a}$, depending on the values of $p_0$ and $p_1$.

In this section we will  discuss only the values of $q_{c,a}$ and the behavior when the barrier can be 
neglected. Values of $q_{c,b}$ and the behavior when the interface actually does interact with the 
barrier will be discussed in Secs. IV and V.

A plot of the numerically determined  contour lines $q_{c,a} = 0,0.1,0.2,...1.0$  is given in Fig.~1. 

\begin{figure}
\begin{center}\hspace* {-1 cm}
\includegraphics[scale=0.35]{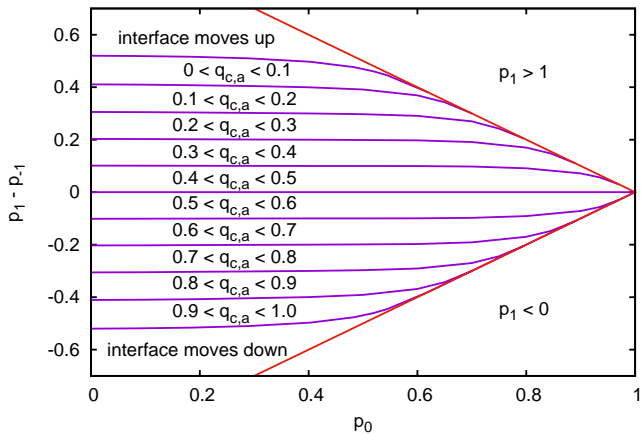}\vglue -1 cm
\end{center}
\caption{Contour lines for the threshold values $q_{c,a} = 0,0.1,0.2,...1.0$, plotted with $p_0$ on the 
	$x$-axis vs $p_1-p_{-1}$ on the $y$-axis. The triangular regions at the right are unphysical 
	with $p_1<0$ or $p_1>1$.  }
\label{fig:contour_a}
\end{figure}

The up-down symmetry of this plot follows from the invariance 
\be
    \{p_1,q,n(x,t)\} \leftrightarrow \{p_{-1},1-q,const-n(x,t)\}.     \label{symmetry}
\ee
At the right hand side of Fig.~\ref{fig:contour_a}, each contour line ends tangentially at one of the two 
phase boundaries $|p_1-p_{-1}| = 1-p_0$ at a finite value of $p_0$.

For finite system size $L$ or when the interface is infinitely high above the barrier, we can also 
consider tilted initial conditions, 
\be 
     n(x,0)= {\rm const} +a x.
\ee
In this case, the speed in general depends on the slope $a$. Thus also 
the critical values of $q$ depend in general on $a$. An exception is the central line $p_1 = p{-1}, q=1/2$ 
in Fig.~\ref{fig:contour_a}. Because of the symmetry Eq.~(\ref{symmetry}), the average speed there is zero,
and the model is thus critical for all slopes. As a consequence, we expect the roughness at criticality, 
and the speed near criticality, to scale according to the EW class along the central line, but according to 
the KPZ class everywhere else. This is indeed observed in simulations, although the scenario is somewhat
delicate.

\begin{figure}
\begin{centering}
\hspace* {-1 cm}
\includegraphics[scale=0.35]{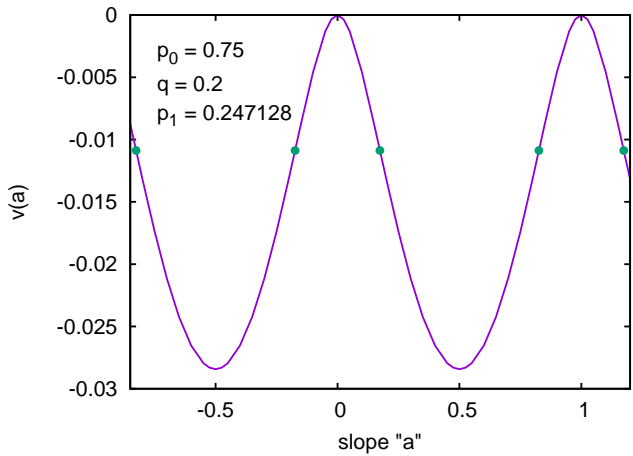}
\vglue -1 cm
\par\end{centering}
\caption{\label{speed_a}  Average asymptotic interface velocity for tilted interfaces
   with tilt $a$. Parameters are such that interfaces without tilt are critical,  $v(0)=0$.
   The curve resembles a cosine, but is clearly distinct from it on closer inspection. The filled dots 
   indicate inflection points where the curvature is zero.} 
\end{figure}

Typical values of the velocity $v(a)$ of the interface, for a fixed set of control parameters, 
are plotted in Fig.~\ref{speed_a} versus the tilt $a$. We see that $v(a)$ is
periodic with period 1. This is indeed easy to prove exactly \cite{Grass-KPZ}: The deterministic 
part of Eq.~(1) gives a speed that is independent of $a$, and the noise is invariant under
a change $a \to a+1$. This periodic dependence of the speed on the tilt is rather unusual for
models in the KPZ universality class, and is responsible for  the slow convergence discussed in the 
following (a more thorough discussion is given in \cite{Grass-KPZ}).

In particular, the periodicity of $v(a)$ implies the existence of inflection points in 
Fig.~\ref{speed_a} where the curvature vanishes. Such points have been discussed previously in the 
context of KPZ \cite{Devillard}. As expected, at such points we also find EW scaling. But we find 
also deviations from the standard KPZ scenario when the curvature $v''(a)$ is small -- either  we are close by an inflection point, or when $q$ is very close to 1/2. As 
the curvature tends to zero, we find that the average interface velocity at large but finite $t$ 
scales as \cite{Grass-KPZ}
\be
     v(t,a) \approx v(a) + {\rm const}~ /\sqrt{t}    \label{vt}
\ee
which is different from both the KPZ behavior where 

\be
     v(t,a) \approx v(a) + const /t^{2/3},    \label{vKPZ}
\ee
and from EW, where there is no power-behaved correction to $v(t,a) \to v(a)$ at all. We conjecture that 
Eq.~(\ref{vt}) holds exactly in the double limit $(v''(a)\to 0, t\to\infty$,  in a not yet precisely 
determined time range $t_1< t < t_2$, where $t_1$ and $t_2$ depend on $v''(a)$.  Our data do not rule out the possibility that $t_2$ is infinite in a finite range of $q$ near $q_c$.

   Because of the slow convergence of the velocity of the interface to its asymptotic value, it becomes difficult 
to pin down precisely the critical parameters where 
$v=0$ and the precise values of $q_{c,b}$.
 These complications do not seem to affect the scenario developed in the 
following and do not prevent us from reaching clear conclusions in the next sections.

\section{Interfaces touching the barrier and finite seeds: Lateral spreading}

\begin{figure}
\begin{centering}\hspace* {-1 cm}
\includegraphics[scale=0.35]{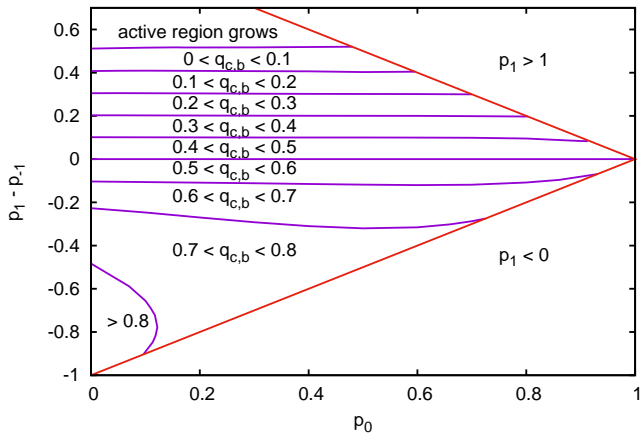}
\vglue -1 cm
\end{centering}
\caption{  Contour lines $q_{c,b} = 0,0.1,0.2,...1.0$ (critical values for spreading from finite seeds), plotted 
with $p_0$ on the $x$-axis vs $p_1-p_{-1}$ on the
$y$-axis.}
\label{fig:contour_b}
\end{figure}

Let us next consider the case where the barrier at $n=0$ is effective, e.g. because the evolution starts from
a finite ``seed". We call the seed ${\cal S}$ of a simulation the set of lattice sites where $n(x,0)>0$,
while $n(x,0)=0$ for all $x$ not in ${\cal S}$. The seed can consist of a single site, of a finite interval,
or of a union of finite intervals. We call the size $R_0$ of the seed the distance between its rightmost and 
leftmost points. Similarly, we denote by $R(t)$ the distance between its rightmost and
leftmost points in the configuration at time $t$ which evolved from this seed, and by $N(t)$ the integrated
height, $N(t) = \sum_x n(x,t)$. We say that a configuration ``survives" up to time $t$, if $N(t)>0$. The 
probability that a configuration survives at least up to $t$ is denoted as $P(t)$. Of course, $R(t), N(t)$,
and $P(t)$ depend on ${\cal S}$. Let us denote as $P_{\infty,{\cal S}}$ the probability that a configuration
starting from ${\cal S}$ survives forever, and $q_{c,b,{\cal S}}(p_0,p_1)$ the largest value of $q$ for which
-- at fixed $(p_0,p_1)$ -- the survival probability $P_{\infty,{\cal S}}$ is zero.

In addition to this survival or extinction transition there is also, in general, a second transition between 
the cases where the average height remains bounded (the interface is attached to the barrier $n=0$) and 
where it detaches ($\langle n(t)\rangle \to \infty$). This transition will be discussed in the
next section. Here we will  discuss only the case where $\langle n(t)\rangle$ stays bounded for $t\to\infty$.

It is clear that configurations can survive forever only, if $N(t)$ and $R(t)$ increase beyond any limit. 
But that suggests also that the detailed form of the seed becomes less and less relevant for the further 
evolution, as time goes on. Thus we expect that the asymptotic scalings of $R,N,$ and $P$ are independent
of ${\cal S}$, and that also $q_{c,b,{\cal S}}(p_0,p_1)$ does not actually depend on ${\cal S}$, as long as 
${\cal S}$ is finite and non-null (an analog situation prevails in DP). This was verified by extensive 
simulations, and the resulting critical values $q_{c,b}$ are plotted in Fig.~\ref{fig:contour_b}
against $p_0$ and $p_1-p_{-1}$. In the following, we shall call $q_{c,b}$ also the 
``critical values for spreading". The extremal points where $n(x,t)>0$ are 
called the left and right ``fronts" of the configuration at time $t$.

\begin{figure}
\begin{centering}\hspace* {-1 cm}
\includegraphics[scale=0.35]{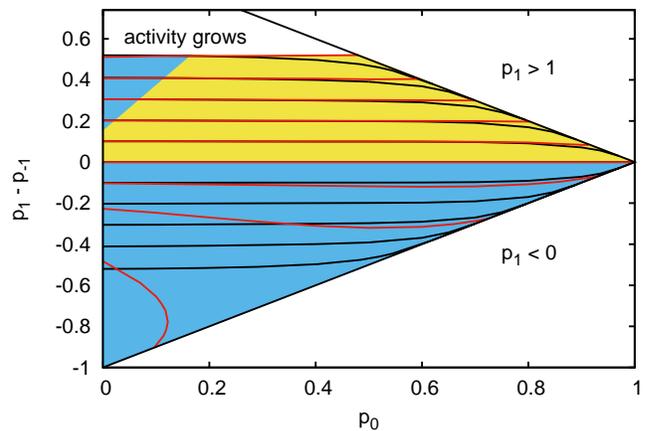}
\vglue -1cm
\end{centering}
\caption{   Both sets of contour lines superimposed, in order to show where the finiteness of the seed enhances
or hinders the proliferation of activity. The region where $q_{c,a} < q_{c,b}$ is colored yellow, while 
the two regions where $q_{c,a} > q_{c,b}$ are colored blue. In the blue regions fronts are pulled and the 
active-dead transition is in the DP class, while fronts are pushed in the yellow region and the active-dead 
	transition is via tent-like interfaces. In the lowest part of the lower blue region (below the 
	lowest curved black line) $q_{c,a}$ is not defined since fronts recede there for all values of $q$ 
	in the absence of a barrier. Thus, in that region the inequality $q_{c,a} > q_{c,b}$ is not applicable,
	but fronts are still pulled there.}
\label{fig:contour_ab}
\end{figure}

\subsection{Pulled versus pushed fronts}

Both sets of contour lines are plotted in Fig.~\ref{fig:contour_ab}, where we also indicated by different colors
the regions where $q_{c,a} > q_{c,b}$ and $q_{c,a} < q_{c,b}$. For reasons that will become clear soon, we call
the former ``pulled", and the latter ``pushed". There is one connected pushed region, and two pulled regions. 
One boundary between pulled and pushed regions is the central line $q=1/2$. The dynamics on this 
line satisfies detailed balance, and the presence of a lower barrier does not break it. The other boundary, in the 
upper left corner of the plot, is known only numerically.

Figure \ref{fig:contour_ab} shows that the existence of fronts created by the lower barrier can either help or 
hinder the spreading of activity. If $q_{c,b} < q_{c,a}$, then for $q \in [q_{c,b}, q_{c,a}]$ a finite seed will
lead (with non-zero probability) to a configuration that spreads forever laterally, while the height of an interface 
starting from an infinitely extended seed would decrease until it hits the barrier. This means that the front 
{\it pulls} the active configuration. If, on the 
contrary, $q_{c,b} > q_{c,a}$, then the active region shrinks and the activity finally dies for $q$ in between
these two critical points. This means that the front {\it pushes} back the activity until it finally dies, 
although the height of an infinitely extended interface without any fronts would increase forever.

Pulled and pushed interfaces are well known from bistable media \cite{Saarloos}, from wetting phenomena (where
wetting fluids correspond to pulled fronts, and non-wetting fluids lead to pushed fronts) \cite{Starov}, and
from spatially extended chaotic systems \cite{Torcini}.

\subsection{Universality classes} 

There are at least five different universality classes for the critical dynamics of interfaces in contact with 
the barrier.

\begin{figure}
\begin{centering}
\includegraphics[scale=0.33]{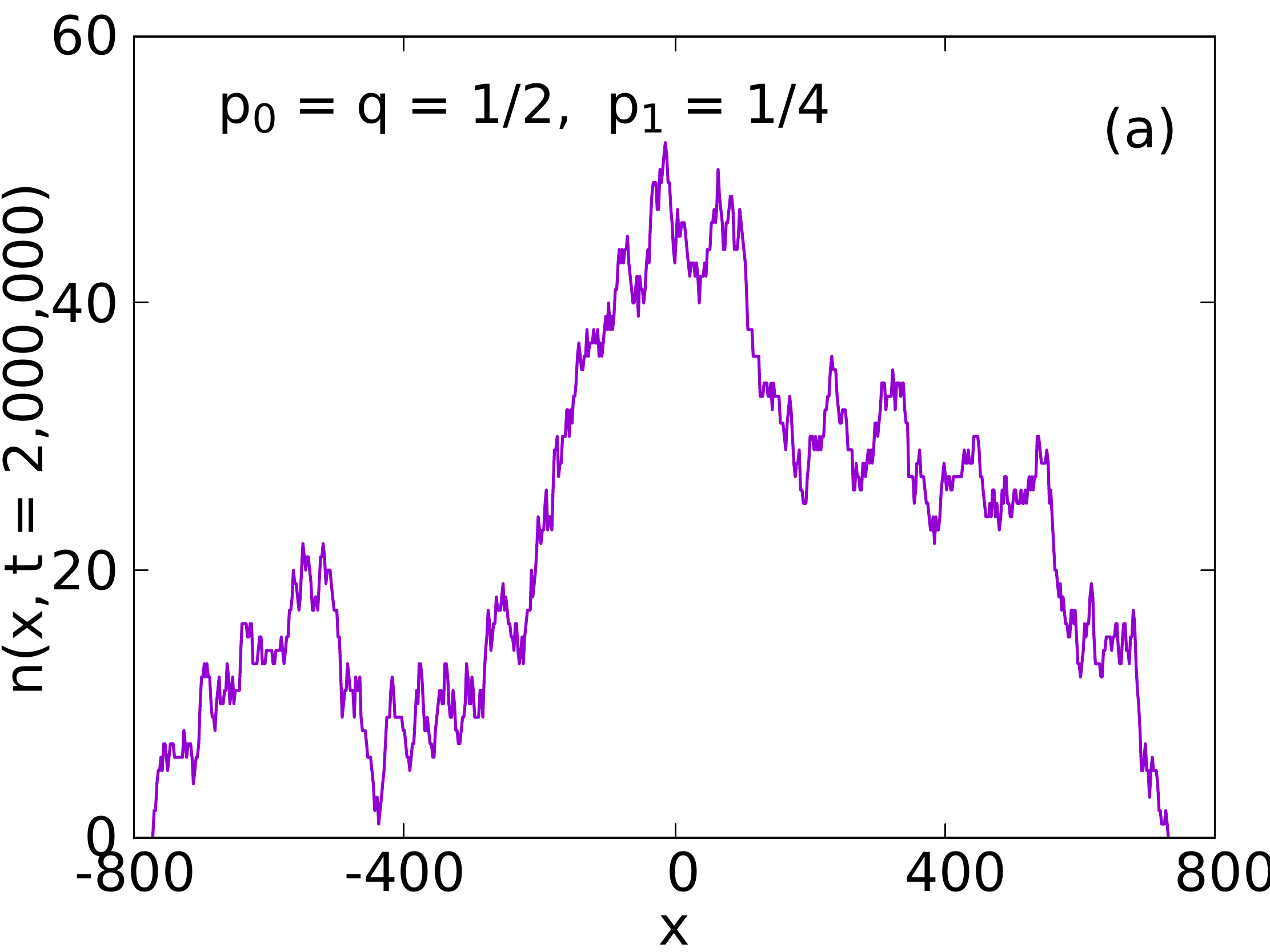}
\includegraphics[scale=0.33]{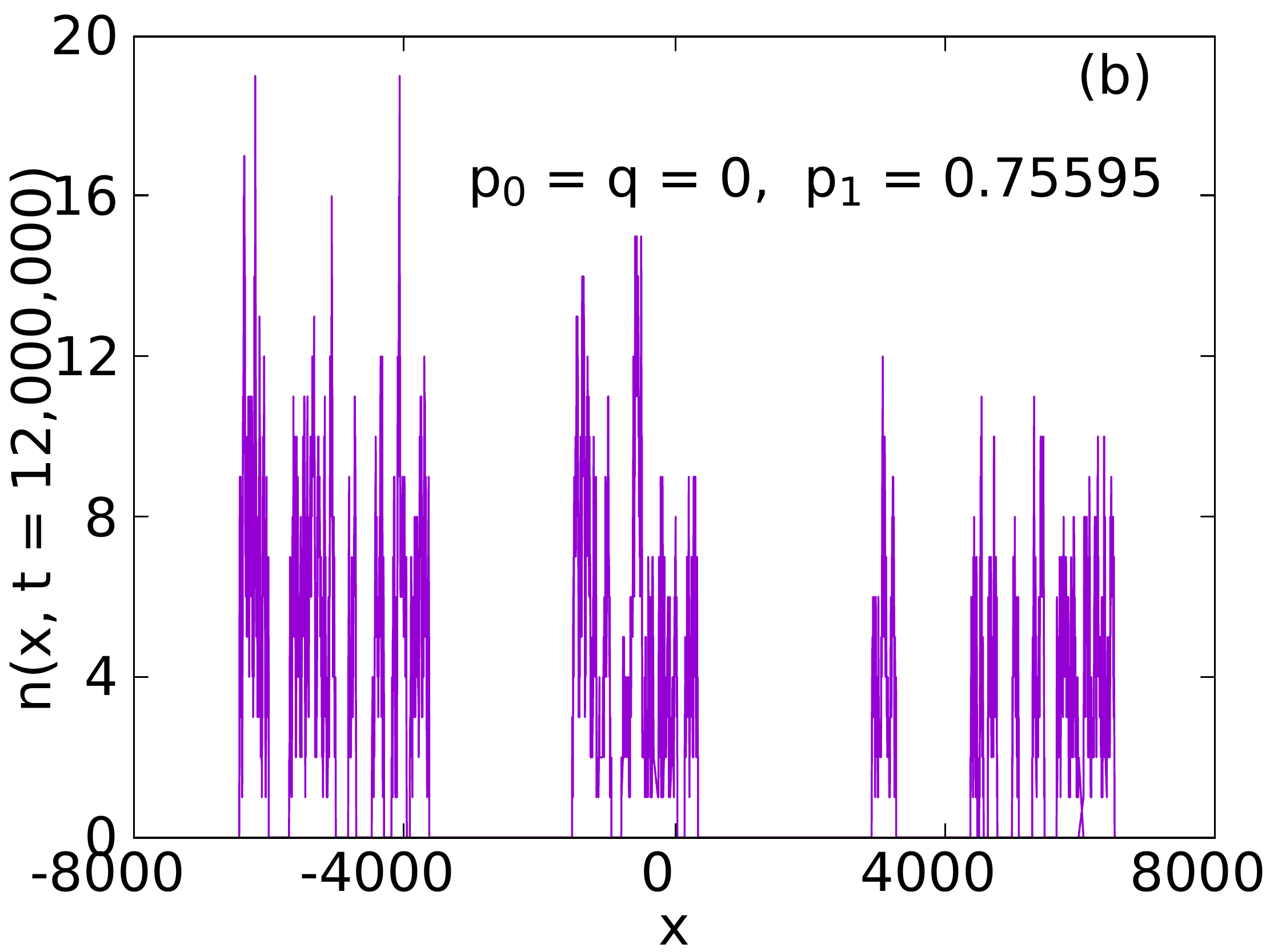}
\includegraphics[scale=0.33]{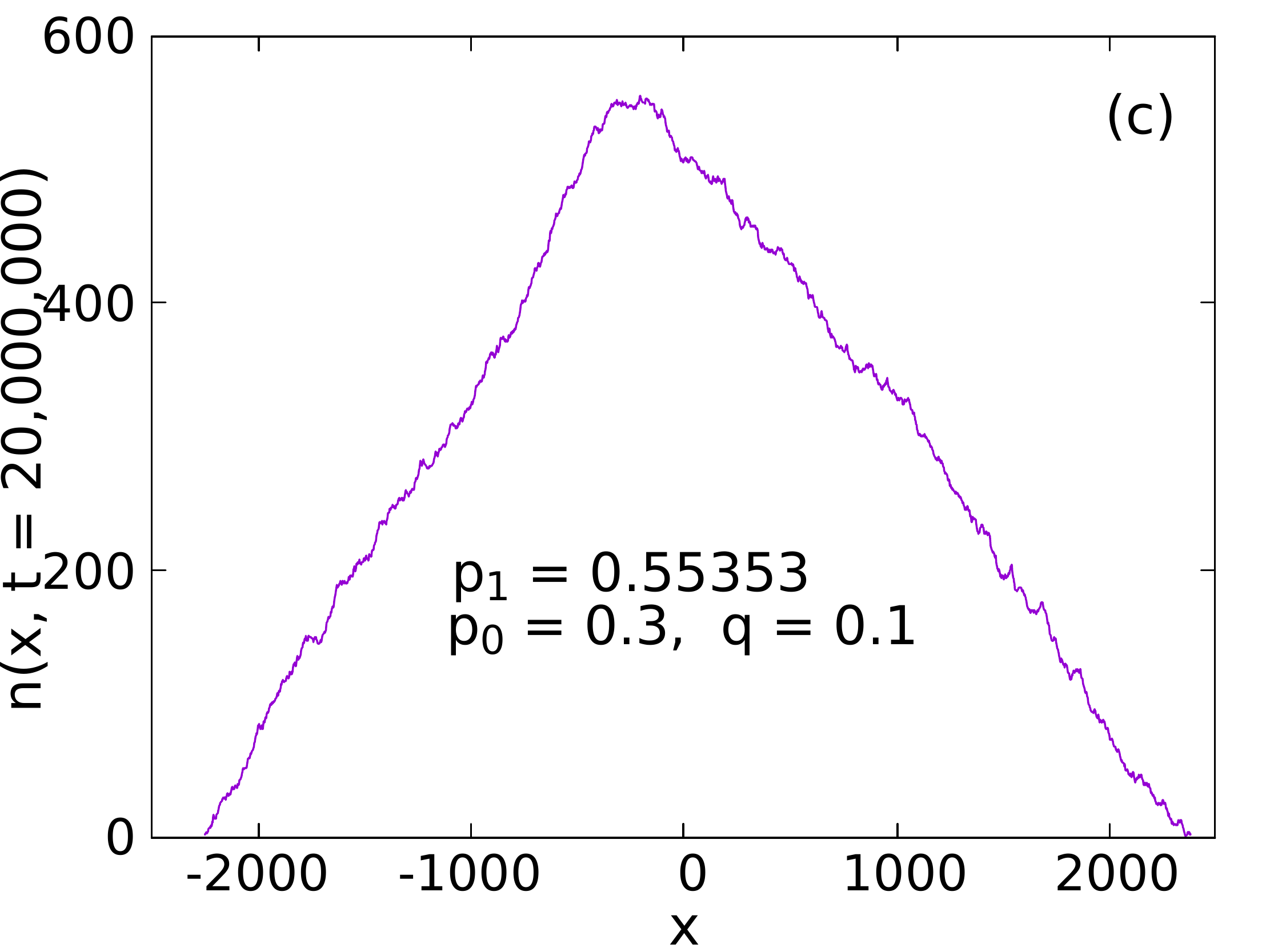}
\end{centering}
\caption{  Typical interfaces obtained at long times when fronts neither push nor pull (a), when they pull 
the active region (b), or when they push it back (c).}
\label{fig:profiles}
\end{figure}

\subsubsection{Compact DP} 

We find compact DP \cite{Essam} for $p_0 = 1$ in addition to $p_1=0$ and $q = 1/2$.
A particularly simple situation prevails when in addition $n(x,0)\leq 1$ for all 
$x$, since in that case $n(x,1)\leq 1$ also for all $t>0$, and the model maps onto the compact DP point in 
the Domany-Kinzel \cite{domany} model (see Sec. 3).

\subsubsection{Clipped EW} 

The next numerically clean and theoretically well understood case happens when infinite
seeds would lead to increasing interfaces in the EW class, and fronts neither pull nor get pushed. This occurs 
when $q = 1/2$ and $p_1 = p_{-1}$. Since EW interfaces are Brownian, an interface between two fronts 
starting from a single point seed is just an arc with $n>0$ clipped out from a Brownian curve, see Fig.~5a. 
Since the fronts neither pull nor get pushed, they perform random walks in $x$. Thus their distance increases as 
$t^{1/2}$, and the average height $\langle n(x,t)\rangle$ of the non-zero part of the interface increases as 
$t^{1/4}$. 
\begin{figure}
\begin{centering}
\includegraphics[scale=0.3]{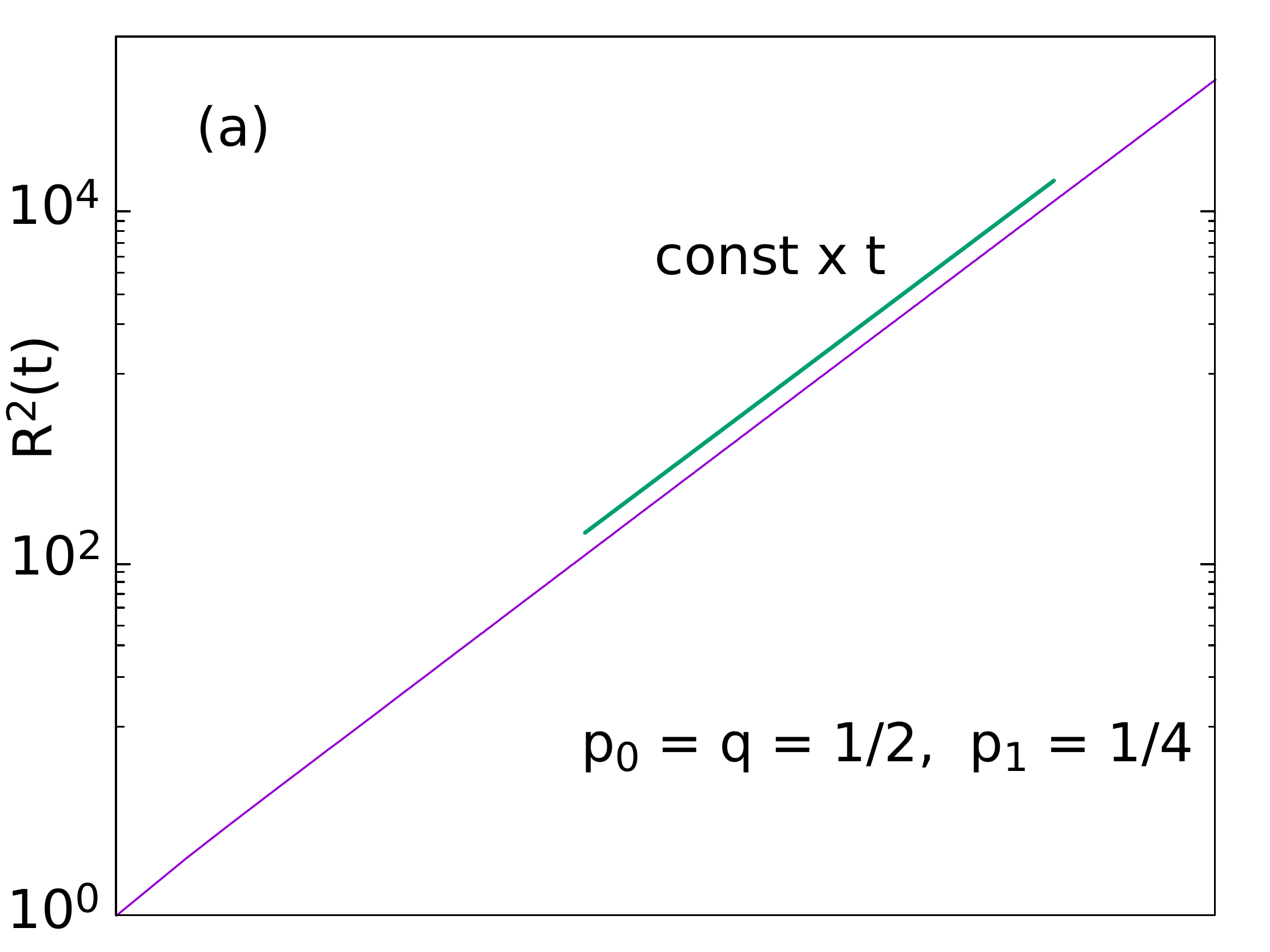}
\includegraphics[scale=0.3]{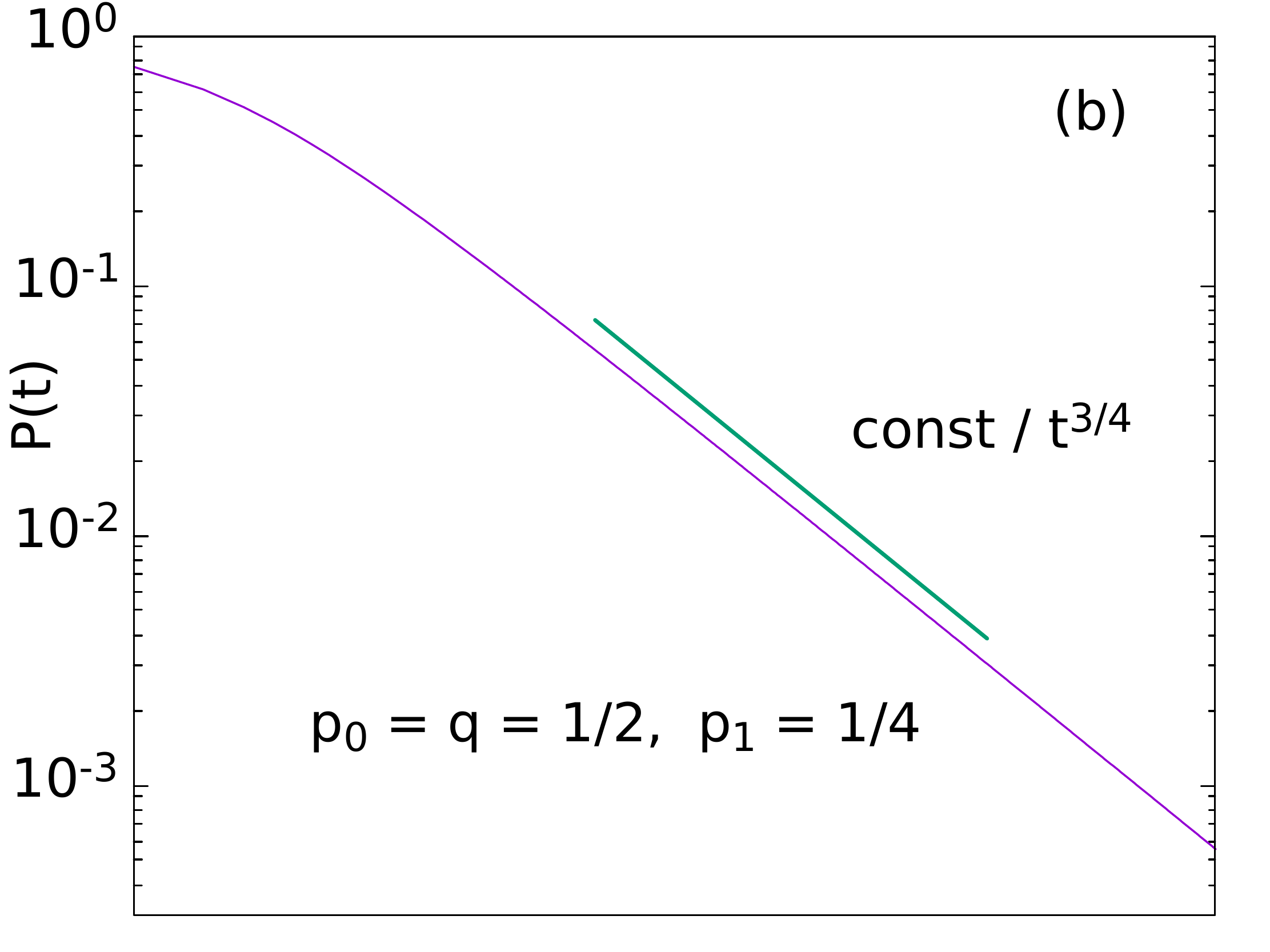}
\includegraphics[scale=0.3]{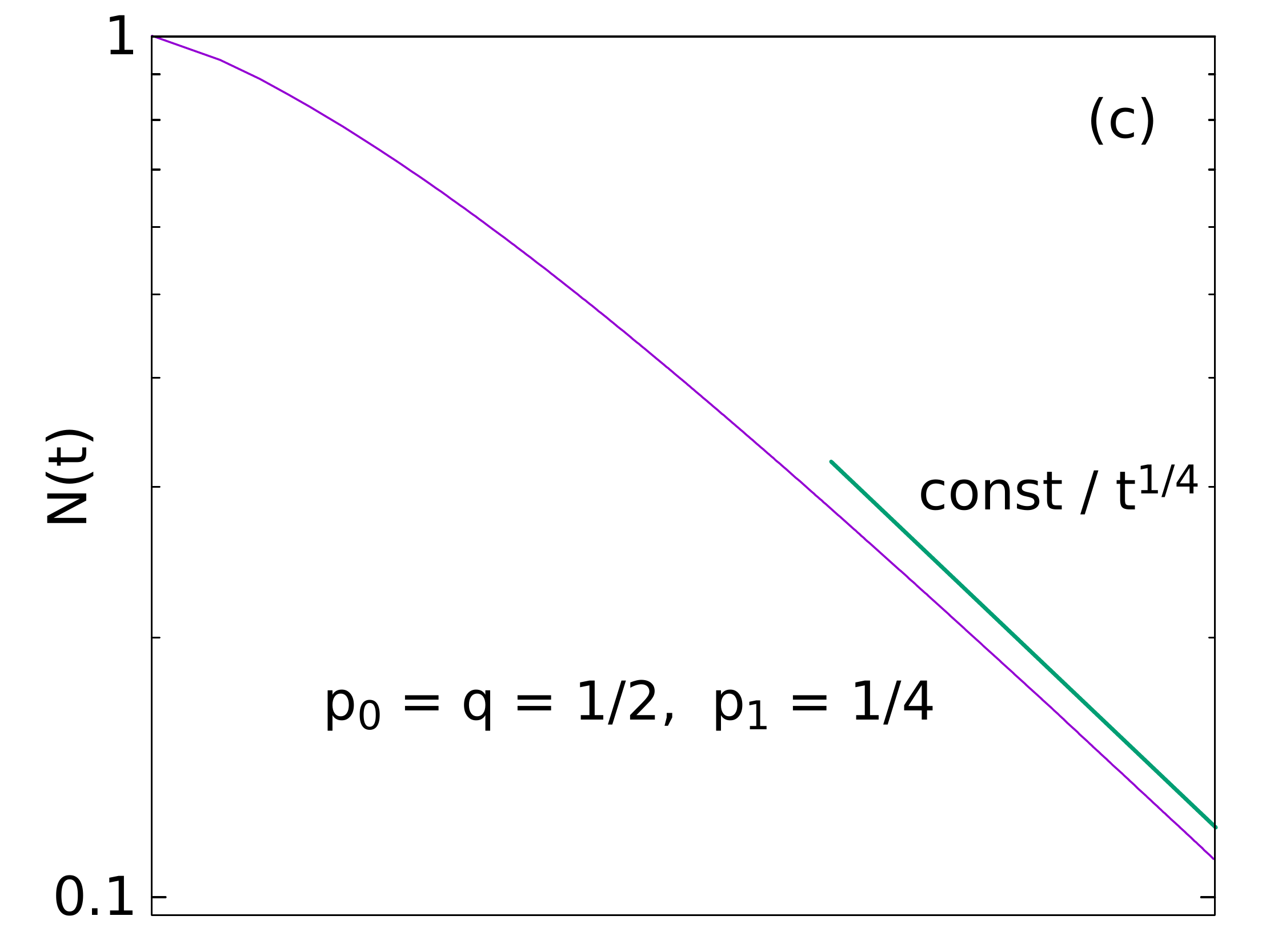}
\includegraphics[scale=0.3]{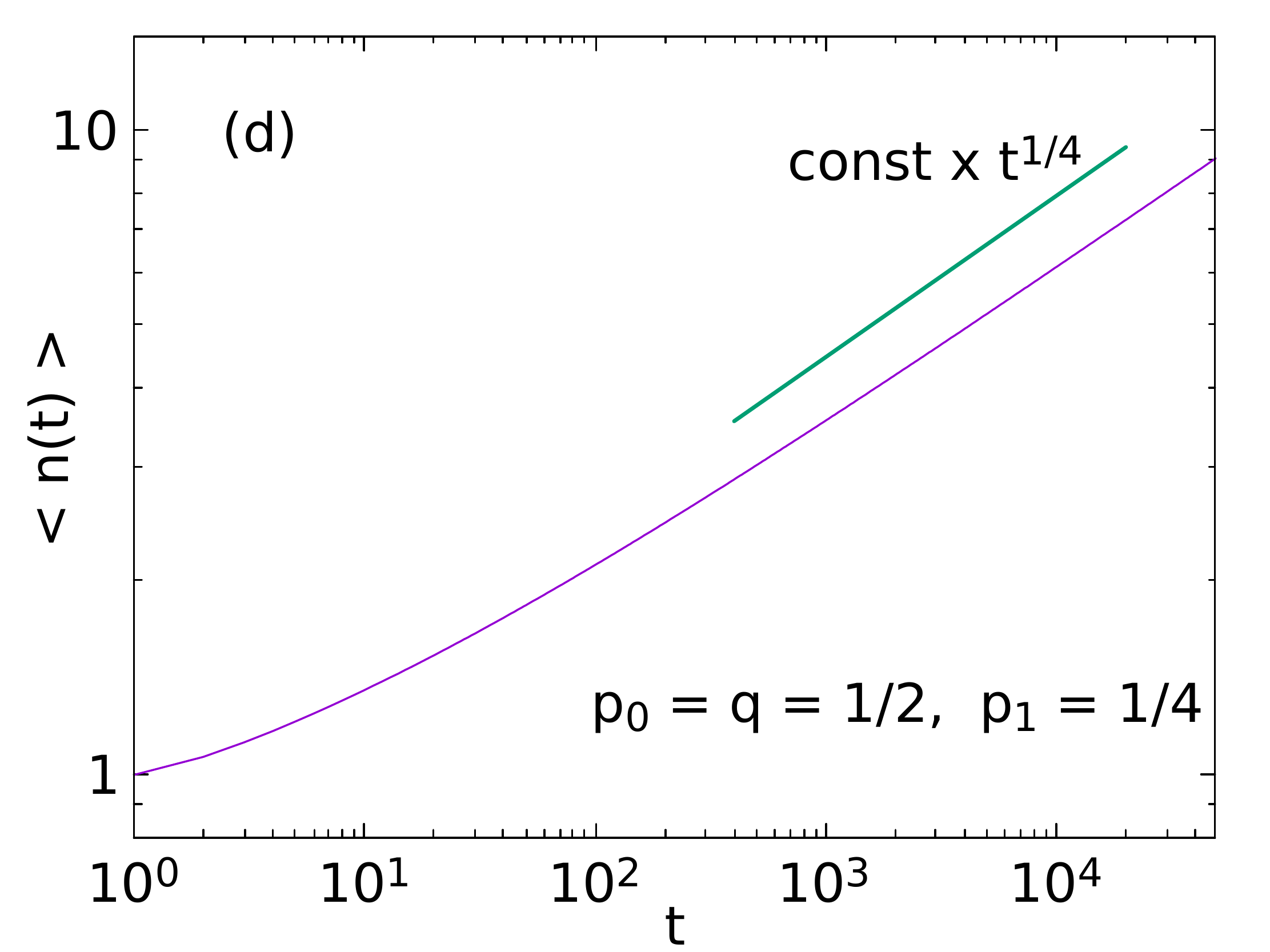}
\par\end{centering}
\caption{\label{fig:EW_clipped}  Log-log plots for time-dependent observables in the clipped EW
   universality class: (a) random mean squared distance of active sites from the origin; (b) probability
   that the interface still moves at time $t$; (c) average number of active sites, i.e. of sites with
   $ n(x,t) > 0$; and  (d) average height $\langle n(x,t) \rangle$ of these sites. In all
   panels, the straight lines indicate the asymptotic scalings, the same set of control 
   parameters is used, and the seed is a single point.}
\end{figure}

If events could die only because the two outer fronts annihilate, we would have $P(t) \sim t^{-1/2}$. But actually
$n(x,t)$ can become zero also inside the arc, creating thereby pairs of (inner) fronts which can then annihilate
later with the outer ones. The result is that actually $P(t) \sim t^{-3/4}$. As a consequence, $N(t)$ does not 
scale as $P(t) \times R(t) \times  \langle n(x,t)\rangle \sim const$ as one would expect naively, but 
$N(t)\sim t^{-1/4}$. As seen from Fig.~\ref{fig:EW_clipped}, all these predictions
are perfectly verified by simulations at the particular point $(p_0,q) = (1/2,1/2)$, but we found the same also
along the entire line $q=1/2$. 

Both compact DP and clipped EW are unstable in the RG sense. When control parameters are perturbed such that they 
stay on the critical surface, they remain in their universality classes for short times and  small distances, but
later cross over to the more stable universality classes of DP [Fig. 5( b)] or the tent phase [ Fig. 5(c)], discussed in 
the next 
paragraphs. Thus DP and the tent phase are more robust, but for the very same reason they are also more difficult
to analyze numerically.
Unless the control parameters are carefully chosen to minimize cross-over effects  (which would be analogous to  the use of  improved Hamiltonians in equilibrium \cite{improvedH1, improvedH2}, and which would be beyond the scope of this article), the simulations are hampered by very slow convergence  due to the presence of nearby clipped EW fixed point.

\begin{figure}
\begin{centering}
\includegraphics[scale=0.3]{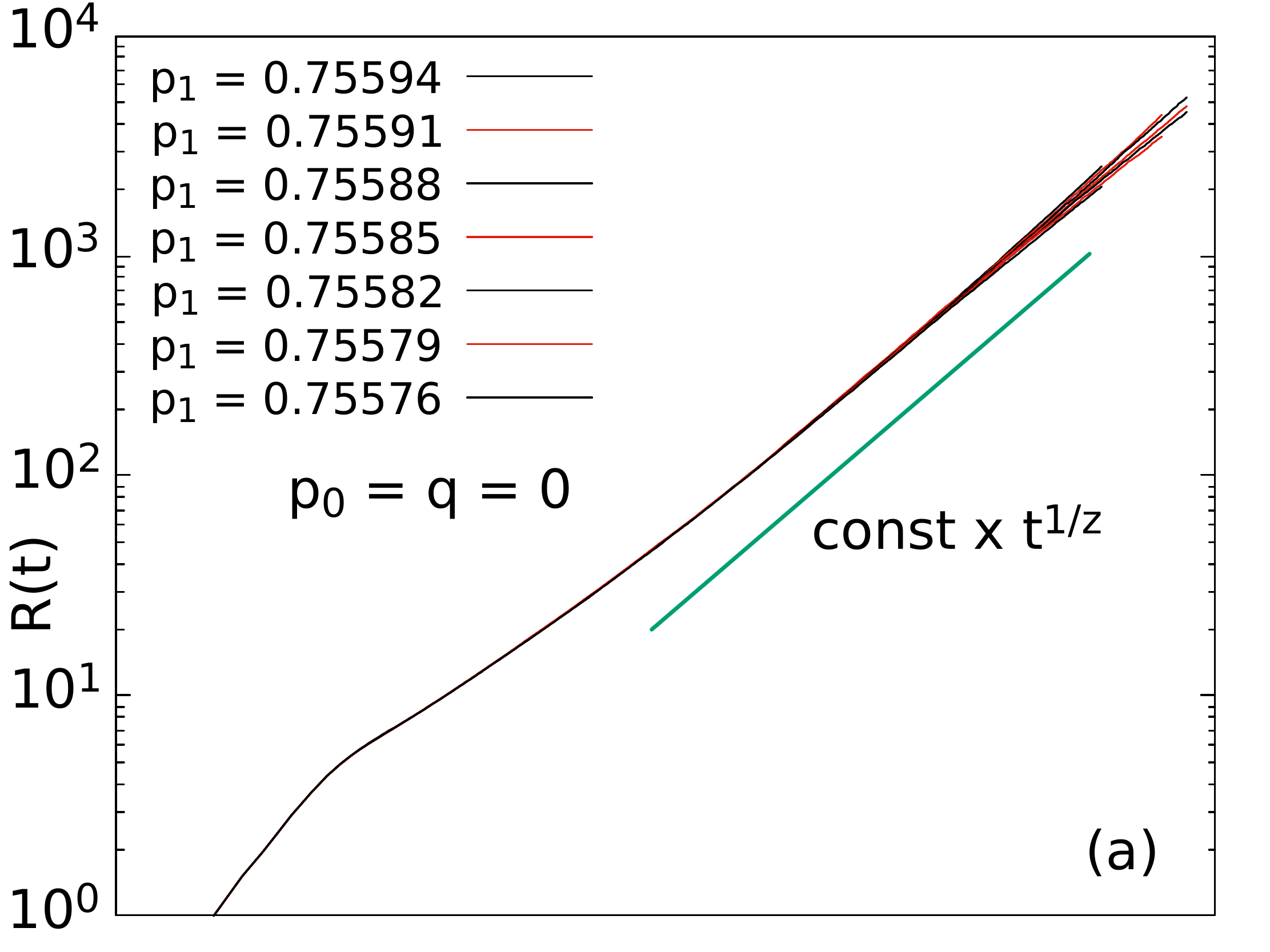}
\includegraphics[scale=0.3]{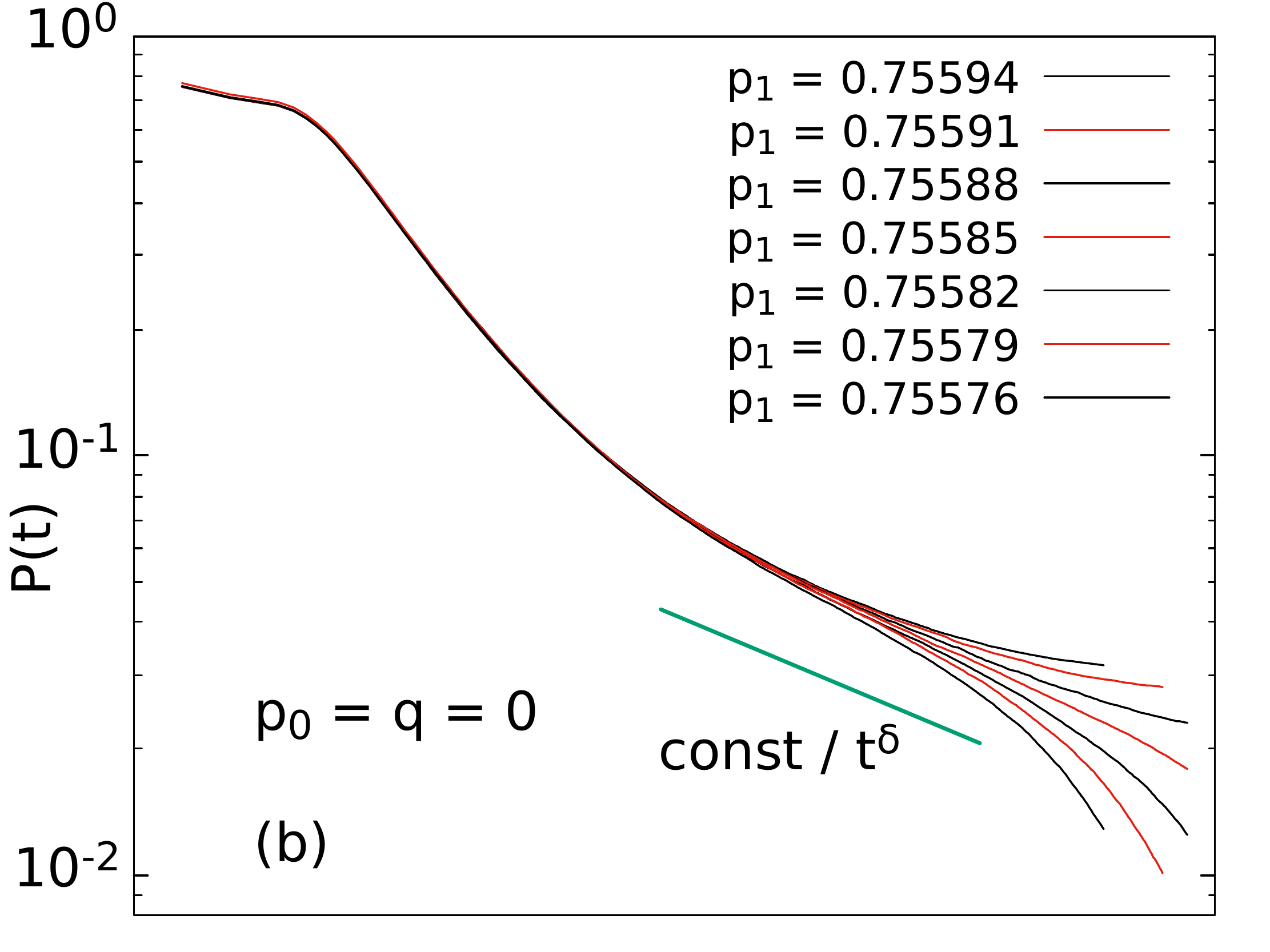}
\includegraphics[scale=0.3]{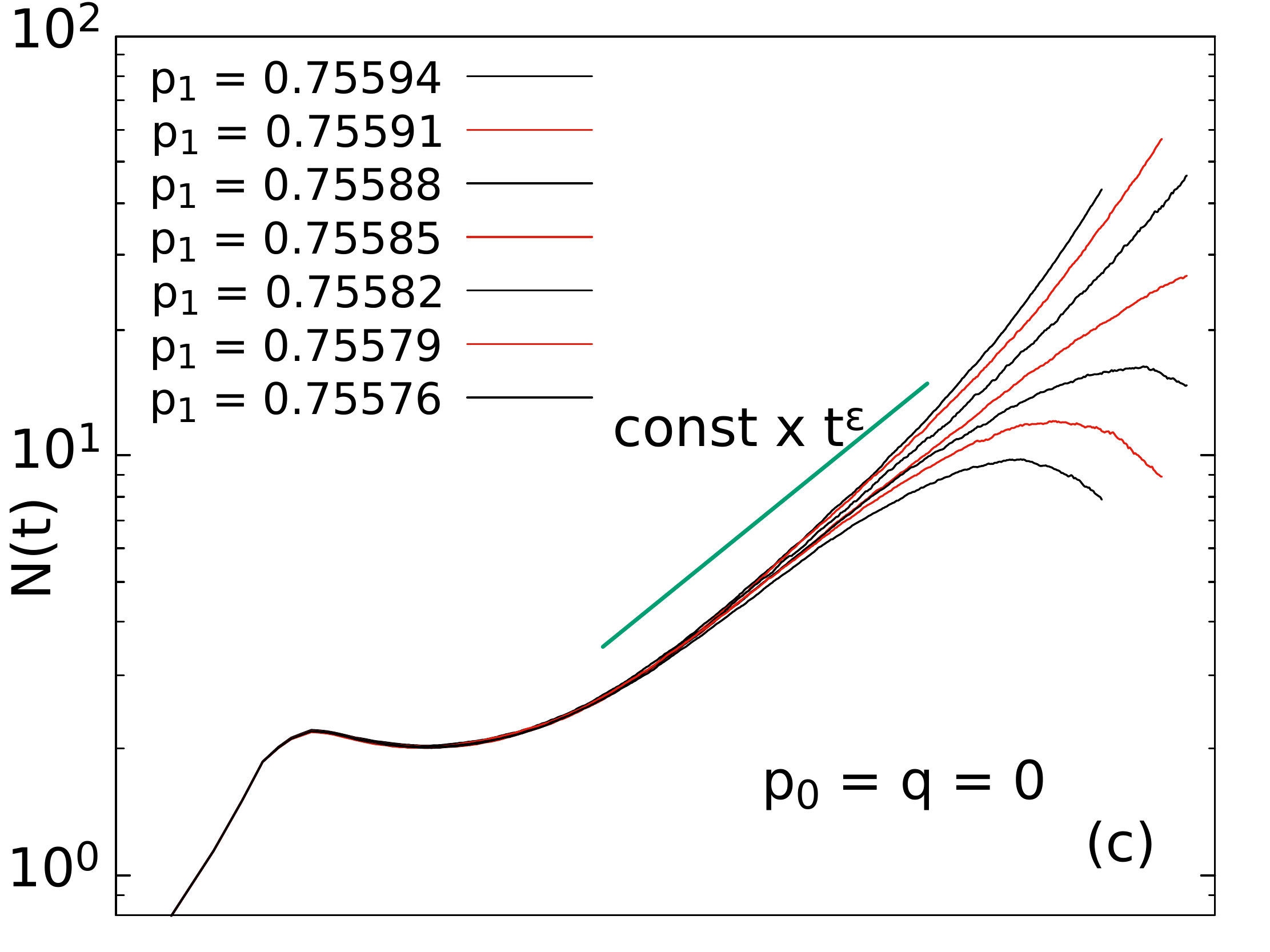}
\includegraphics[scale=0.3]{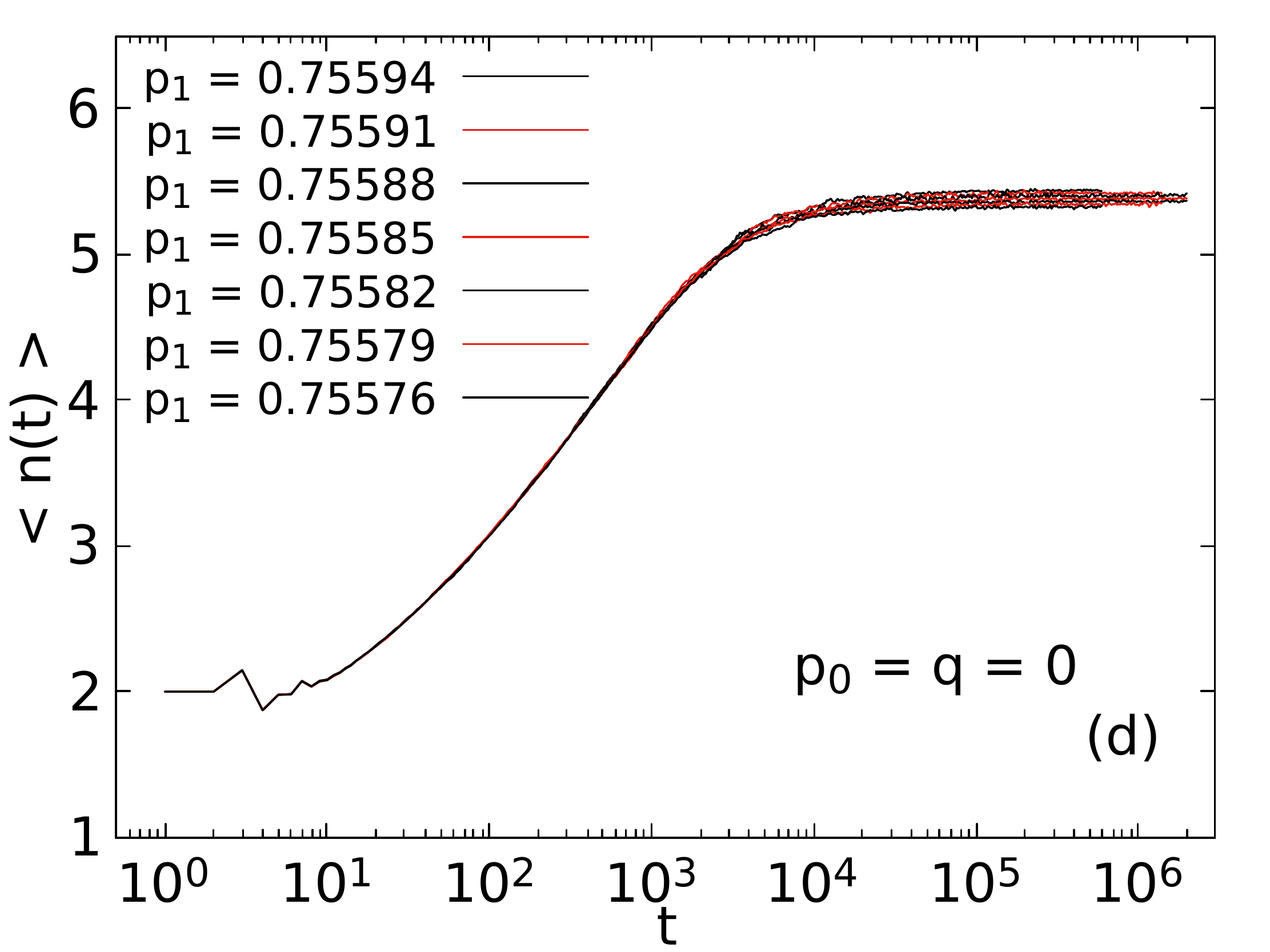}
\par\end{centering}
\caption{\label{fig.0-0}  The same four observables as in the previous figure, but now for
   a set of parameter values in the DP universality class. Again a point seed was used. In all panels 
   (except panel (d)) the straight lines indicate the asymptotic scaling expected for DP \cite{Jensen}. 
   All panels show log-log plots, except for panel (d) which is log-linear. Notice that 
   $\langle n(x,t) \rangle =1$ for ordinary DP.}
\end{figure}

\begin{figure}
\begin{centering}
\includegraphics[scale=0.35]{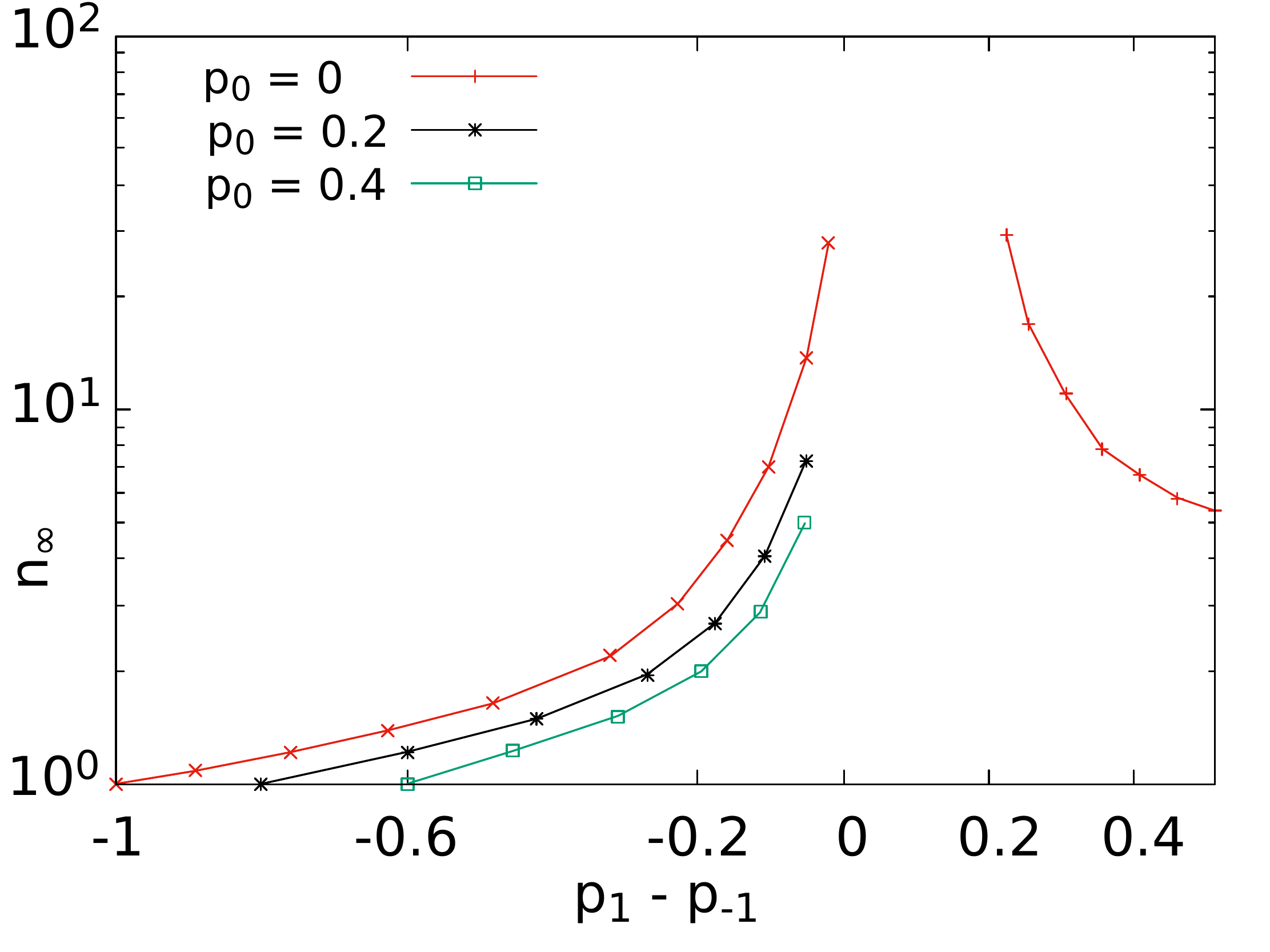}
\vglue -.5cm
\par\end{centering}
\caption{\label{fig:DP-n_infty}  Log-linear plot of asymptotic average interface heights of 
	active sites, at three lines on the critical DP transition manifold.} 
\end{figure}

\begin{figure}
\begin{centering}
\includegraphics[scale=0.3]{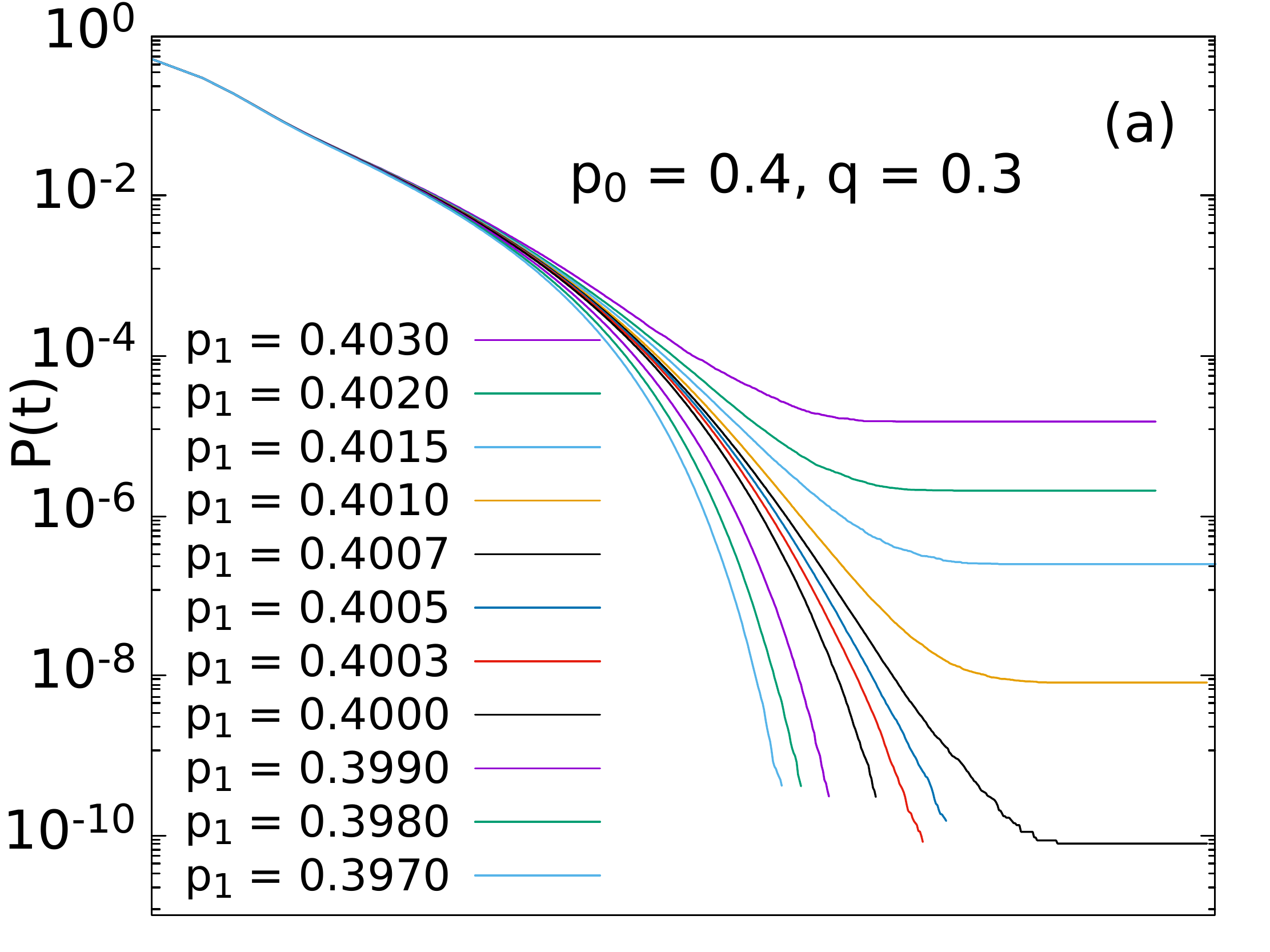}
\includegraphics[scale=0.3]{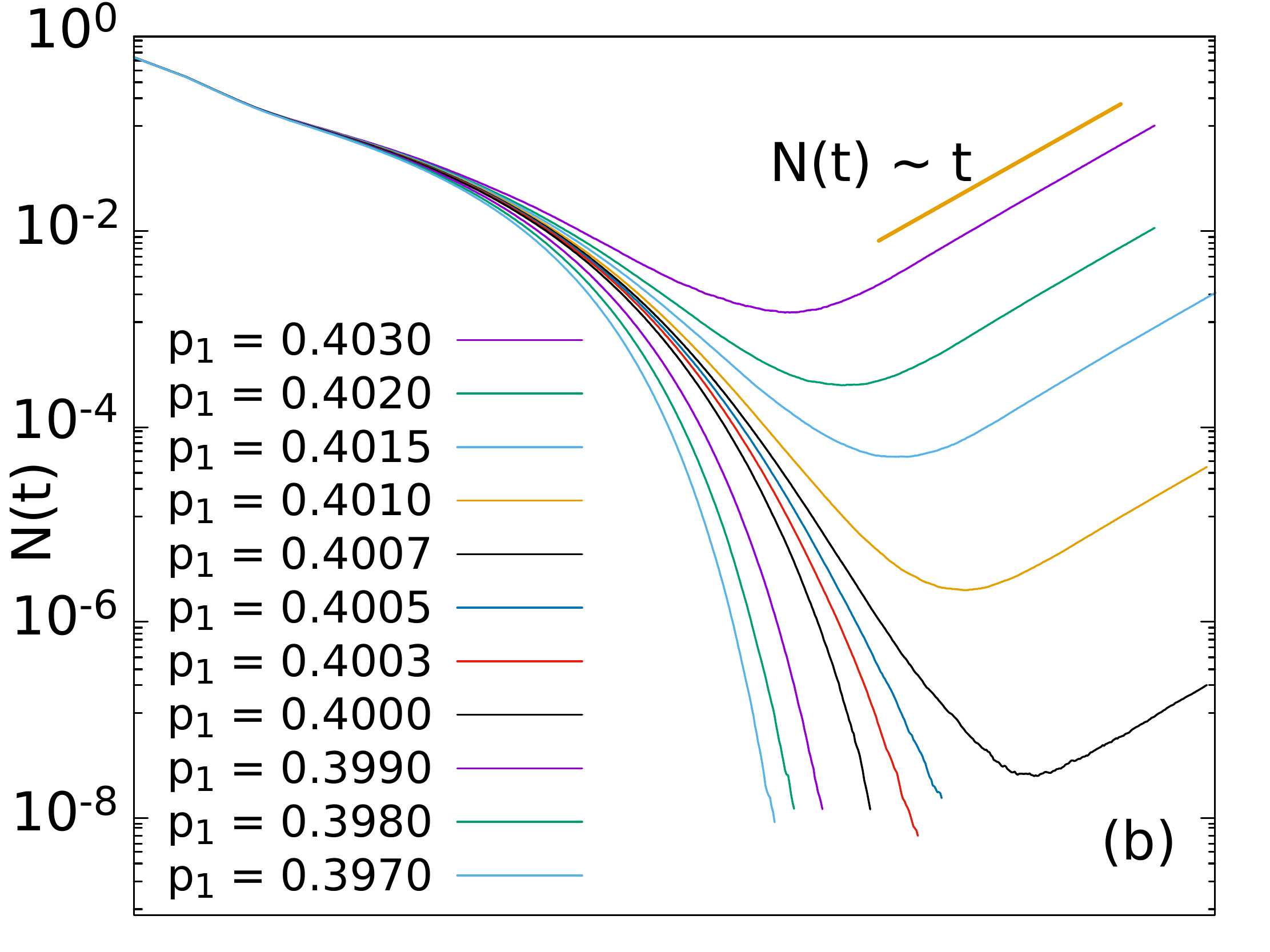}
\includegraphics[scale=0.3]{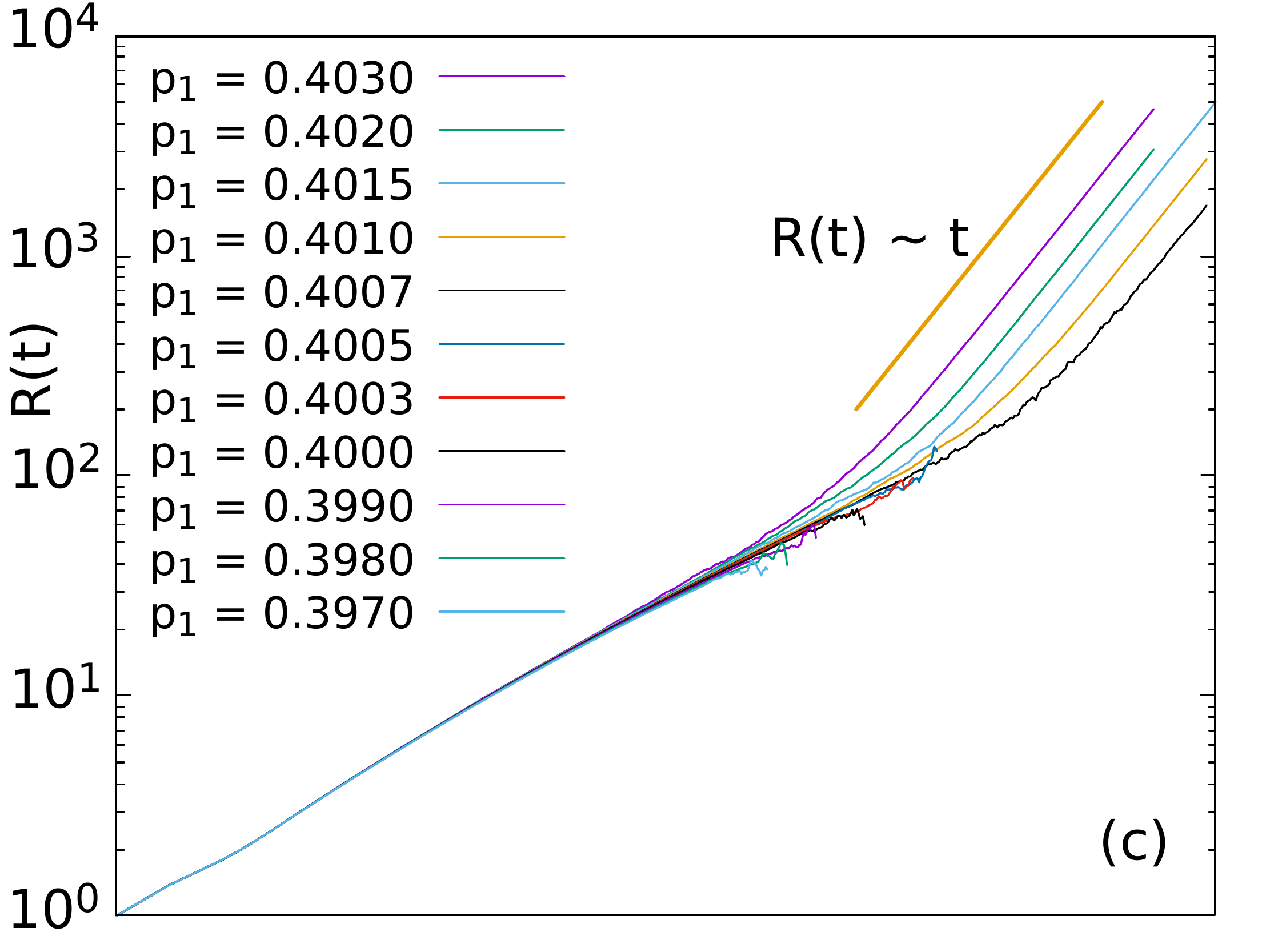}
\includegraphics[scale=0.31]{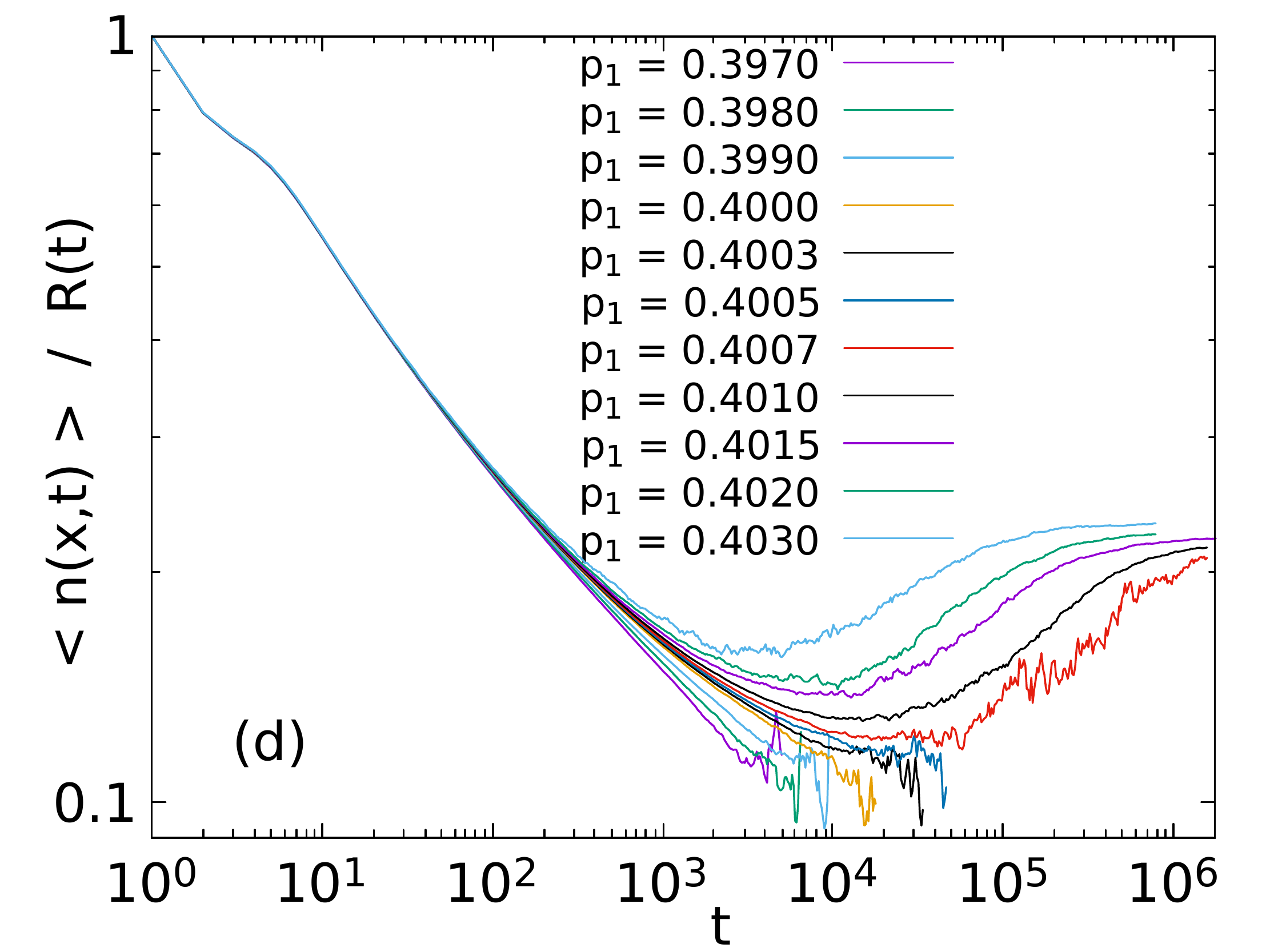}
\par\end{centering}
	\caption{\label{fig:EW-.4-.3}  Panels (a) to (c) show log-log plots of the same observables as 
	in the first three panels of Fig.~\ref{fig.0-0}, but now for $p_0=0.4$ and $q=0.3$ where
	the transition is of tent type. Again a point seed was used. In panels (b) and (c) the 
	straight lines indicate the asymptotic scaling, which corresponds just to a linear expansion. Panel (d)
	shows the ratio between the average height $\langle n(x,t)\rangle$ and the average lateral size $R(t)$.}
\end{figure}

\subsubsection{Directed percolation} 

The behavior is particularly simple when $p_1=0$ and when the initial state has $n(x,0)\leq 1$ for all
sites $x$.  Then it is easily seen that $n(x,t) \in \{0,1\}$ also for all later times. In this case we
are left with two independent control parameters $q$ and $p_0$, and the model can be mapped 
exactly onto the well known Domany-Kinzel (DK) cellular automaton model \cite{domany}. In the DK model,
$n(x,t)=1$ means that site $x$ is infected at time $t$, while $n(x,t)=0$ means that it is susceptible. 
Neighboring pairs ``01" and ``10" at positions $x\pm 1$ infect site $x$ at the next time step with probability
$p$, while pairs ``11" infect it with probability $p'$. In the present
notation, $p=q$ and $p'=p_0$. If $q=p_0$, then two infected neighbors have the same chance to infect the
site between them as does a single infected neighbor. This corresponds to site DP. On the other hand,
if the second infected neighbor has the same chance to infect it as the first one (provided the first
neighbor was not successful), we have bond DP and $p_0=2q-q^2$. Finally, when $p=1/2$, the critical behavior
of the DK model is in the compact DP \cite{Essam} class, which is thus also the universality class of our
model for $p_1=0, q=1/2$, and $n(x,0)\leq 1$ for all sites $x$.

For arbitrary finite seeds,  DP occurs whenever the front is pulled.  This is the case 
along the boundary line $p_1=0$, but it also occurs in both blue regions of Fig.~\ref{fig:contour_ab}. Numerical 
results obtained at $p_0=q=0$ for the same four observables as in the previous figure are shown in Fig.~\ref{fig.0-0}. 
Here and in the following examples we do not determine the critical $q$ for fixed $p_0$ and $p_1$, but we fix 
$p_0$ and $q$ and then determine $p_{1,c}$.
We see clearly that the known DP critical exponents \cite{Jensen} hold for $p_1 =0.75587(1)$, although the scaling 
sets in rather late -- as was expected from the fact that there can be slow cross-overs from other unstable fixed
points. Similar results were found in all regions where we expect DP scaling. Notice, however, that the 
activity per site (the value of $n(x,t)$) can become
arbitrarily large. More precisely, at the critical surface $\langle n(x,t)\rangle$ tends to a finite value 
$n_\infty(p_0,p_1)$. For a few selected values of the control parameters, $n_\infty(p_0,p_1)$ is plotted in 
Fig.~\ref{fig:DP-n_infty}. As $p_1 \to 0$ we have $n_\infty(p_0,p_1)\to 1$, while $n_\infty(p_0,p_1)$ diverges 
when $p_{-1}-p_1 \to 0$, as in this limit the behavior has to cross over to clipped EW. But it diverges also
when the other boundary between pulled and pushed phases (see Fig.~3) is approached.

\subsubsection{The tent transition and the tent phase} 

A scaling very different from DP is observed when fronts try to push the activity back, 
but are pushed out by the bulk. This transition
between spreading and non-spreading phases has many features of a first order transition, although we will avoid 
this terminology which, after all, is rooted in equilibrium theory.

Plots like those in Figs.~\ref{fig:EW_clipped} and \ref{fig.0-0}, but for a typical point on the tent transition
manifold, are shown in Fig.~\ref{fig:EW-.4-.3}.
The first striking observation is that $P(t)$ does not seem to follow a power law at the critical point. Rather, 
the system seems to go through a bottle neck as $t$ increases and one is very slightly overcritical. More precisely,
$P(t)$ seems to decrease faster than with a power law, until the very few surviving ``clusters" have reached a critical
size, after which they grow linearly and $P(t)$ no longer decreases at all. This qualitative behavior is typical of
cluster growths near critical points in first order transitions in various systems like nucleation \cite{Kashchiev},
magnetism \cite{ball}, wetting \cite{cahn}, and co-infections \cite{coinfect}. This interpretation is confirmed 
by the other panels of Fig.~\ref{fig:EW-.4-.3}. Figures \ref{fig:EW-.4-.3}(b) and \ref{fig:EW-.4-.3}(c) show that both $N(t)$ and $R(t)$ grow linearly with 
$t$, as soon as the bottleneck is passed. Moreover, a closer analysis shows that $N(t)/P(t) = R(t)$ up to finite 
(cluster) size corrections, when $q>q_{c,b}$. This is precisely what we expect, if the active region is compact
without any holes. Finally, Fig. \ref{fig:EW-.4-.3} (d) shows that the ratio between the lateral size $R(t)$ and the  average
$\langle n(x,t)\rangle$ of  the surviving clusters tends to a constant, as expected from   the  tent-like shape of the asymptotic curve in Fig. \ref{fig:profiles}, independent of $p$. What changes with the distance from the critical point is the speed of expansion. Panels Figures \ref{fig:EW-.4-.3}(b) and \ref{fig:EW-.4-.3}(c) show that
this speed increases linearly with $q$. 

At first it might seem that it is impossible to determine precisely the transition point from simulations
like those presented in Fig.~\ref{fig:EW-.4-.3}. The curves for $P(t)$ indicate that $p_{1,c} \approx 0.4$ to $0.4007$,
but not more.
But there is an easy way to obtain more precise estimates. First of all, Fig.~\ref{fig:EW-.4-.3} was, just like 
Figs.~\ref{fig:EW_clipped} and \ref{fig.0-0}, for point seeds. Using extended and tent-shaped seeds with $R_0$ 
up to several thousands, we can avoid the bottleneck, since for such large seeds $P(t)\approx 1$ even very close to 
the transition point. Second, in such runs we can measure rather precisely the asymptotic speed $v_\infty $ 
of (lateral and  vertical) growth. Extrapolating linearly to $v_\infty=0$ gives then the critical point. For the 
present case we get, e.g., $p_{1,c} = 0.40022(1)$.
This is indeed how most of the simulations underlying Fig.~2 were performed.

From Fig.~\ref{fig:EW-.4-.3}(d) we also see that $\langle n(x,t)\rangle /R(t)$ tends, for $t\to\infty$ and for all 
values of $p_1$ to the same constant, 
\be
   \lim_{t\to\infty} \langle n(x,t)\rangle /R(t) = 0.26(1).
\ee
If, indeed, the shapes becomes triangular in this limit, then this is the asymptotic slope of the triangles. 
The fact that this limit is approached from below is due to statistical fluctuations which tend to round off
the top of the triangles for small $t$ and for $p_1$ very close to $p_{1,c}$.

Figure \ref{fig:EW-.4-.3} (and many simulations at other control parameter values) confirms that
surviving clusters at late times 
are tent shaped; see Fig.~4(c). This is easily understood. Since interfaces without fronts would be in the KPZ
universality class, the velocity of tilted interfaces would depend on their tilt, such that flat interfaces
move fastest. At the same time, since fronts are pushed out by the bulk and the bulk is pushed back by the 
fronts, statistical fluctuations at one of the front 
positions cannot propagate back into the interior of the cluster. Together these two observations imply that 
statistical fluctuations propagate outward, as long as the local tilt is larger than a critical value. It is this 
critical tilt which then defines the slope of the triangular profiles that evolve. More precisely, let us denote 
the triangle slope as $a_\Delta$. Then we have
\be
   q_{c,a}(a_\Delta) = q_{c,b}.
\ee
This was indeed confirmed, within measurement errors, for all tested control parameters. In particular, it was
verified that $q_{c,a}(0.26) = 0.3$ for $p_0=0.4$ and $p_1 = 0.40022$, within the statistical errors.

A further important difference between the tent and DP transitions is that $\langle n(x,t)\rangle$ diverges 
(linearly) for $t\to\infty$ as soon as one has passed the tent transition while, as we had seen, it converges 
to a finite value
in slightly supercritical DP. It is only at a finite distance above the DP transition that interfaces detach,
in a second transition, from the barrier. That second transition will be discussed in Sec.~V.

\subsubsection{Clipped KPZ} 

The next universality class is that when fronts neither are pushed nor pull, and interfaces without
fronts would be in the KPZ universality class. This is realized along the short approximately straight line in
Fig.~3 that cuts off the triangle in the upper left corner. In this case most of the remarks about capped EW
should apply {\it mutatis mutandis.} But numerical verification was impossible because of the problems discussed
in \cite{Grass-KPZ}.

\section {Detachment transitions of pulled interfaces from the barrier}

Let us finally discuss the critical behavior(s) at $q = q_{c,a}$, in the case where $q_{c,b} < q_{c,a}$ 
and where therefore fronts are pulled. In this case we have interfaces attached to the barrier for 
$q_{c,b} < q < q_{c,a}$, while they must be detached for $q \gg q_{c,a}$ \footnote{Notice that the 
following discussion applies only to the part of phase space where $q_{c,a}$ is defined, thus it does 
not apply at the region below the lowest curve in Fig.~\ref{fig:contour_a}. In that region, interfaces 
can never detach from the barrier.}. Clearly, the {\it detachment transition} is exactly 
at $q_{\rm detach} = q_{c,a}$. Otherwise (if $q_{\rm detach} > q_{c,a}$) we would have bistability in 
the interval $q_{c,a} < q < q_{\rm detach}$: An interface which started at height $n_0 \gg 1$ would 
be detached forever, while one starting at finite $n_0$ would remain attached forever. An even 
stranger situation would prevail if $q_{\rm detach}$ depends on $n_0$. In the following we shall 
present numerical evidence that seems to exclude such exotic behavior, and which strongly suggests
that $q_{\rm detach} = q_{c,a}$.

As seen from Fig.~\ref{fig:contour_ab}, there are two disjoint regions (both indicated in blue) where 
$q_{c,b} < q_{c,a}$. In the larger (lower) region, $q_{c,a} >1/2$ and thus  
\be
    \langle n(x,t)\rangle \sim n_\infty +c\; t^{\alpha},    \label{alpha}
\ee
with $c>0$ (it does not really matter whether $\alpha=1/3$ as for proper KPZ or $\alpha\approx1/2$, as seen 
effectively for $q\approx 1/2$ \cite{Grass-KPZ}). On the other hand, we have $c<0$ in the smaller (lower) 
blue region in Fig.~\ref{fig:contour_ab}. In the first case a critical interface would move towards the 
barrier for small $t$, while it would move away from it in the second case. We shall call the first 
``uneasy detachment", while the second is ``easy detachment". We should expect that the effect of the 
barrier is different in both cases, and that we have indeed two different universality classes.

In order to decide this and to obtain critical exponents, we turn again to simulations. We choose values of 
$p_0$ and $p_1$ in the upper or lower blue regions of Fig.~\ref{fig:contour_ab}, and depending on them we 
choose $q \approx q_{c,a}$. We start each run with $n(x,t) = n_0 \geq 1$ for all even $x$. Although we 
made simulations also for $n_0>1$, we show  results only for $n_0=1$. As observables we 
measure the average and variance of the height $n(x,t)$ as a function of $t$, and the height distribution 
$\rho(n,t)$ at fixed $t$. In particular we look at $\rho(0,t)$, i.e. the probability that the interface touches
the barrier at any given site. In all simulations we used $L = 2^{19}$. 

Although we looked also at other control parameters, we show  results only for the two choices, \\

 A: $(p_0,q) = (0.5,0.7)$, where $p_{1,c} \approx 0.15$; and \\
\indent B: $(p_0,q) = (0.1,0.0)$, where $p_{1,c} \approx 0.71$. \\

Notice that we again switched from fixing $p_0$ and $p_1$ to fixing $p_0$ and $q$. As in the previous
subsections this was done purely for numerical convenience, without any deeper reason.
Choice A corresponds to uneasy detachment, while detachment is easy for choice B.

\subsection{Uneasy detachment}

\begin{figure}
\begin{centering}
	\includegraphics[scale=0.35]{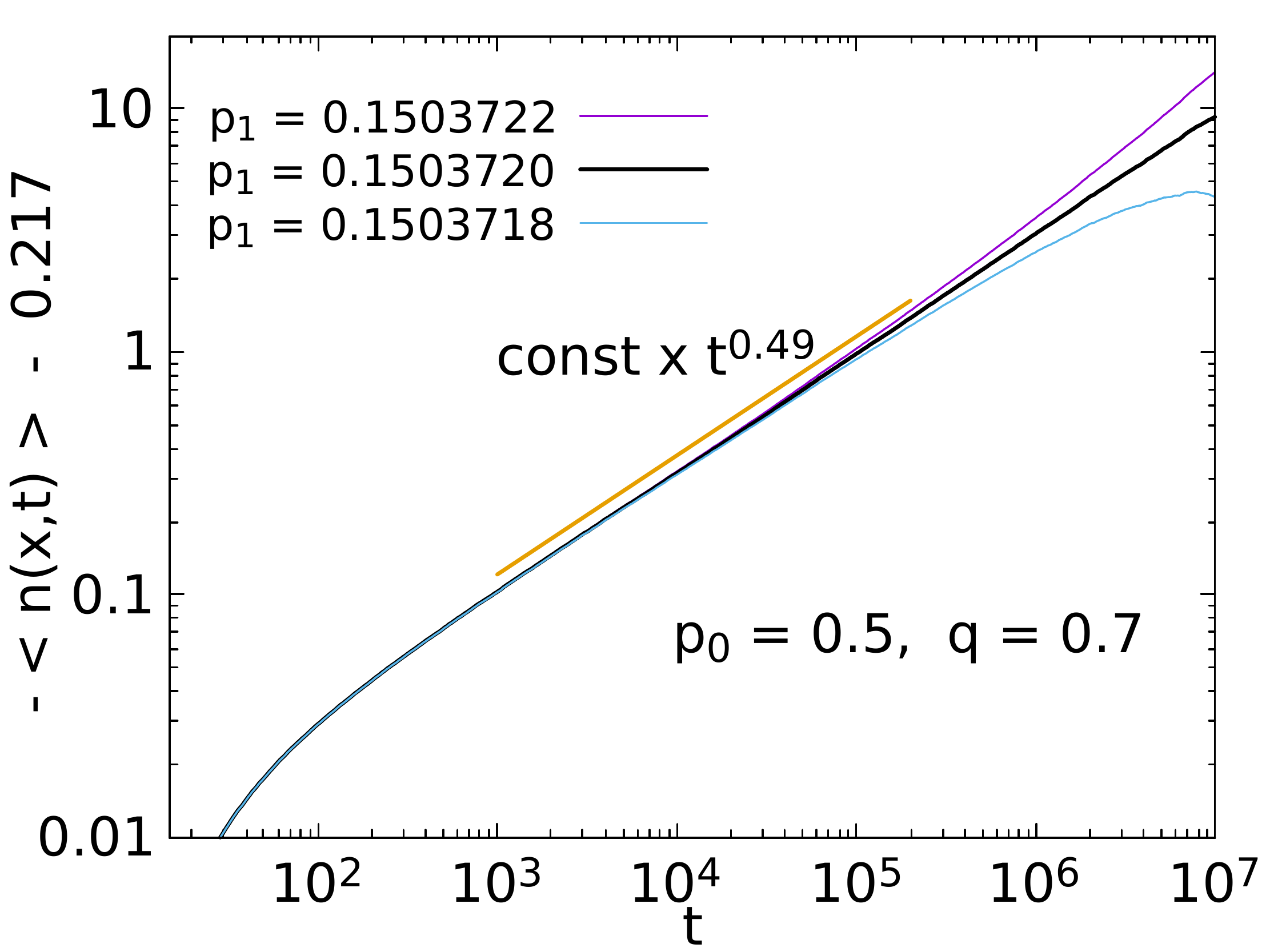}
\par\end{centering}
   \caption{\label{fig:n_A}  Average heights for $p_1 \approx p_{1,c}$ at point A in absence 
   of a barrier, plotted against $t$. More precisely, a log-log plot is shown of minus the average height, 
   shifted by an amount which is chosen such that the curve for $p_1 = p_{1,c}$ is straight for the largest 
   range.}
\end{figure}

\begin{figure}
\begin{centering}
        \includegraphics[scale=0.35]{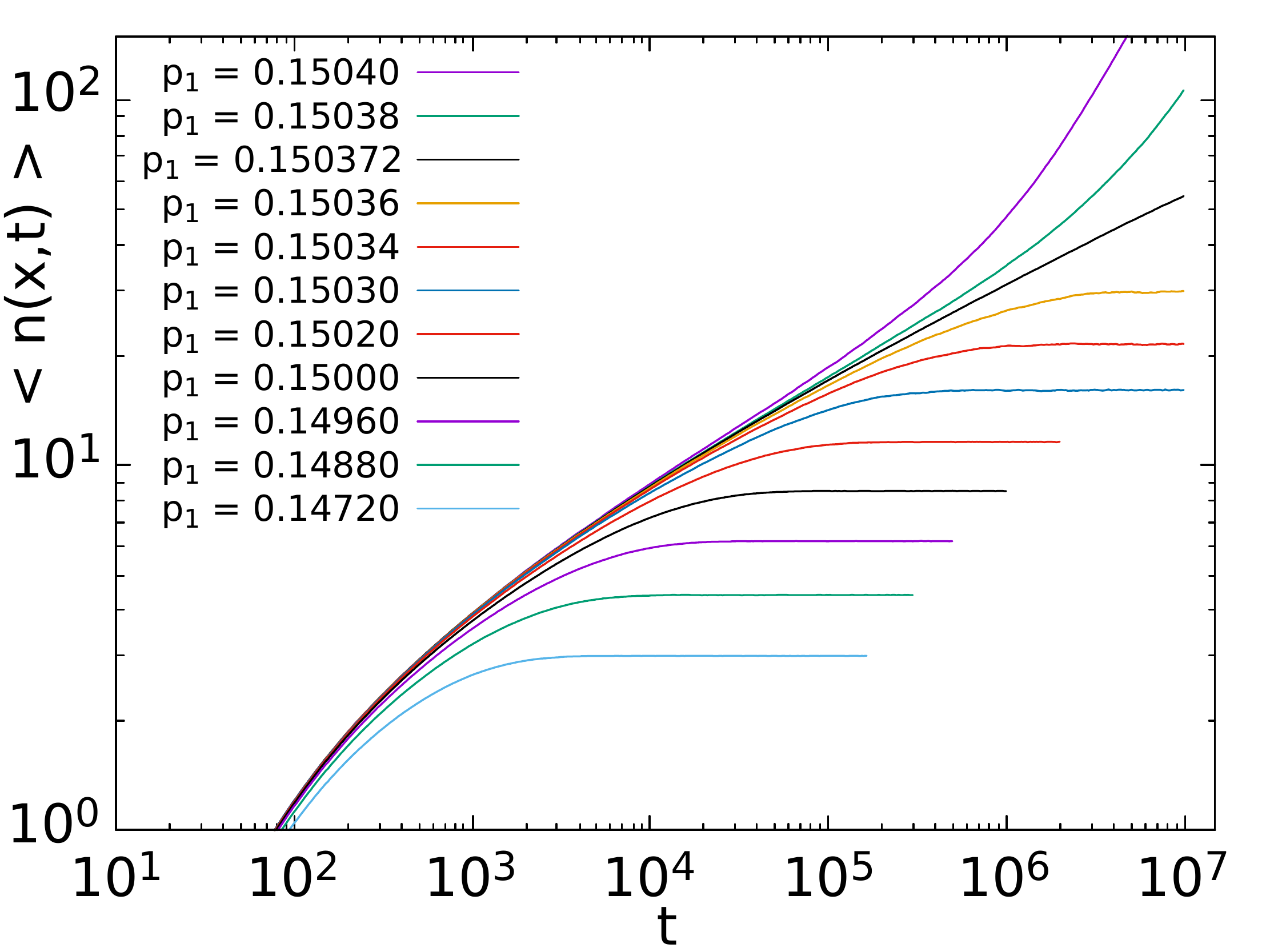}
\par\end{centering}
        \caption{\label{fig:n_t_A}  Average heights for $p_1 \approx p_{1,c}$ at point
	A (uneasy detachment) in the presence of a barrier, plotted against $t$.}
\end{figure}

In order to obtain a precise estimate of $p_{1,c}$ at point A, we show in Fig.~\ref{fig:n_A} plots of the 
average interface height versus $t$ in absence of the barrier. More precisely, we show log-log plots of 
$c -\langle n(x,t)\rangle$, where $c=-0.217$ is chosen such as to maximize the region where the critical
curve shows a clean power law. We see indeed a perfect power law, with exponent  $\alpha \approx 0.49$, \cite{footnote2}
for $p_{1,c} = 0.15037198(10)$. Notice that the error of $p_{1,c}$ takes into account the uncertainty of $\alpha$ 
and is indeed largely dominated by it.

Results for $\langle n(x,t)\rangle$ versus $t$ in presence of the barrier, at several values of $p_1$, 
are shown in Fig.~\ref{fig:n_t_A}. Although there are huge corrections to it, a conventional finite size 
scaling (FSS) ansatz
\be
   \langle n(x,t)\rangle = t^{-\mu} \Phi[(p_1 - p_{1,c})t^{1/\nu_t}]
\ee
seems to describe the asymptotic behavior. From the scaling of the critical curve we then obtain 
$\mu = 0.23(1)$, while the scaling of $n_\infty$ against $p_1 - p_{1,c}$ gives $\mu\nu_t = 0.34(1)$ and 
thus $\nu_t=1.5(1)$. Notice, however, that these estimates assume that the curvature of the critical curve 
seen in Fig.~\ref{fig:n_t_A} does not continue to much higher values of $t$. If it does continue, then we 
actually cannot exclude that $\mu=0$ and the above scaling laws could be all wrong -- although we would 
consider this as extremely unlikely. 

\begin{figure}
\begin{centering}
        \includegraphics[scale=0.35]{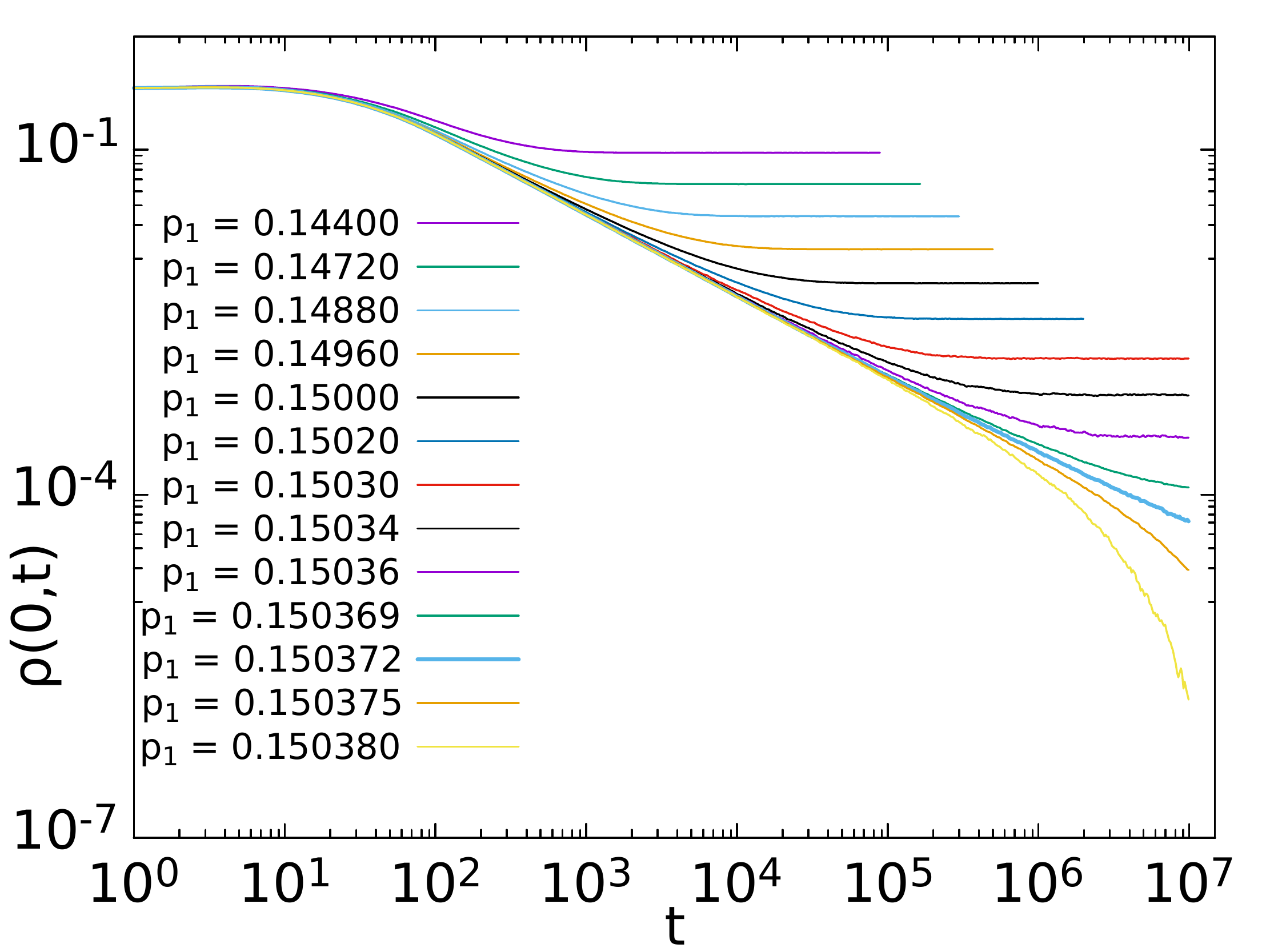}
\par\end{centering}
        \caption{\label{fig:P_0-A}  Fractions of the interface that are at height $n=0$ for 
	several values of $p_1$, at point A and in the presence of a barrier.}
\end{figure}

\begin{figure}
\begin{centering}
        \includegraphics[scale=0.35]{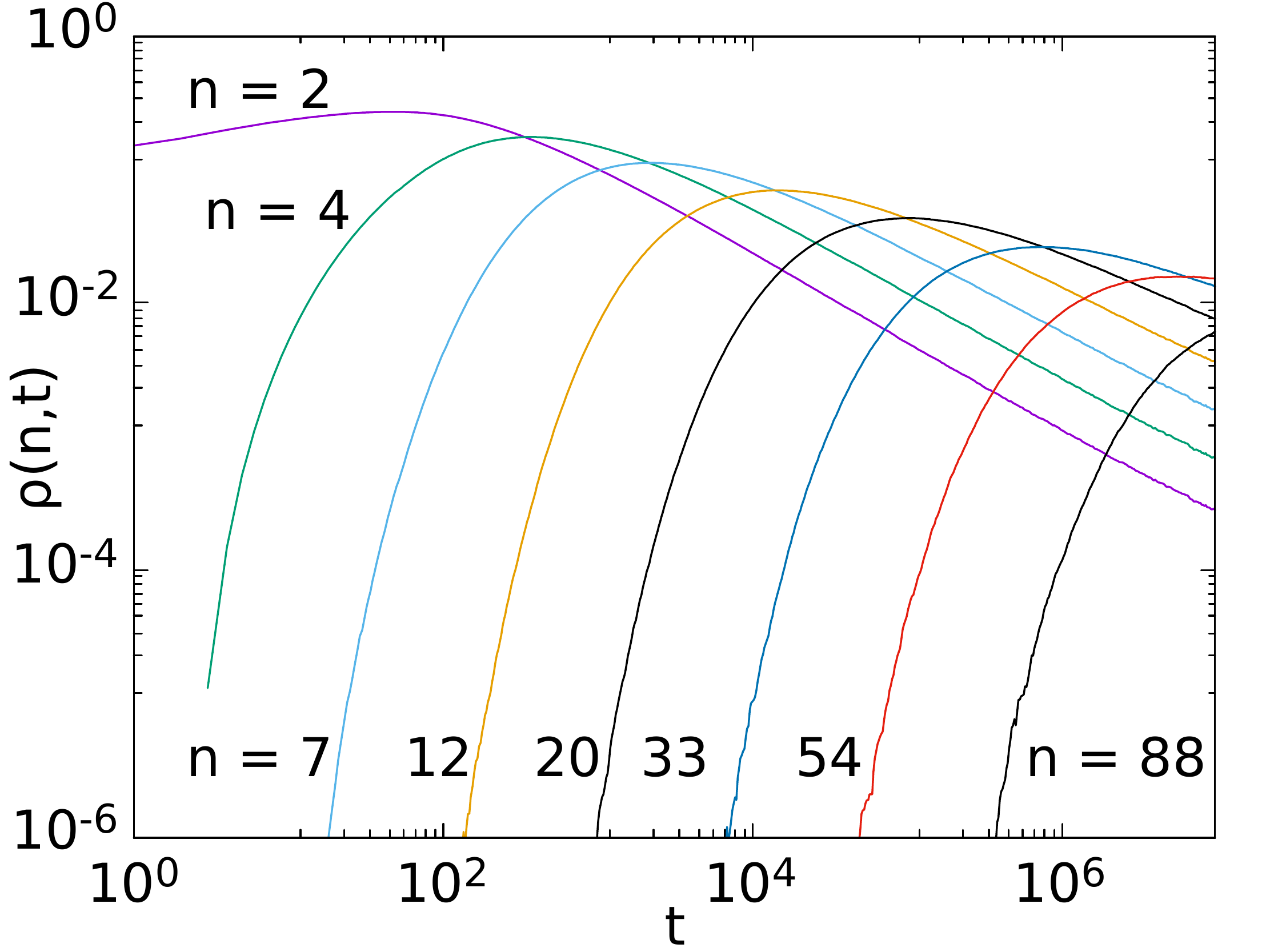}
\par\end{centering}
        \caption{\label{fig:A_crit-rho_nt}  Fractions of the interface that are at heights 
	$n=2,\;4,\;7,\;12 \ldots 88$ for $p_1=p_{1,c}$, at point A and in the presence of a barrier.}
\end{figure}

If there are scaling laws for $\rho(n,t)$, they should be different for $n=0$ and $n>0$. Values of $\rho(0,t)$ 
are plotted in Fig.~\ref{fig:P_0-A} for several values of $p_1$. We verify again that $p_{1,\rm detach}=p_{1,c}$, 
and we see similar FSS as for $\langle n(x,t)\rangle$. But a closer inspection shows that the 
power law at criticality,
\be
   \rho(0,t) \sim t^{-\delta},
\ee
is actually much less clean than suggested by a fit in the large interval $10^2 < t < 10^7$. While that
would suggest $\delta = 0.72(2)$, the data for $t>10^6$ show a clear deviation which leads to $\delta \leq 0.60(3)$.
We cannot, indeed, give any non-zero lower bound on $\delta$ with any confidence. From Fig.~\ref{fig:P_0-A} we 
can also read off values of $\rho_{\infty,0} = \lim_{t\to\infty}\rho(0,t)$ for $p_1 < p_{1,c}$. Again, a casual
analysis would suggest a power law $\rho_{\infty,0} \sim (p_{1,c}-p_1)^y$, but a more careful inspection shows 
that there are so large deviations that we refrain from quoting a value for the exponent $y$.

For each $n\geq 2$, $\rho(n,t)$ has a maximum at a value $t=t_n$ which increases with $n$ (see 
Fig.~\ref{fig:A_crit-rho_nt}). For large $n$, this increase follows roughly a power law, but again corrections 
to it are too large to present a reliable estimate of the exponent.

\subsection{Easy detachment}

\begin{figure}
\begin{centering}
        \includegraphics[scale=0.35]{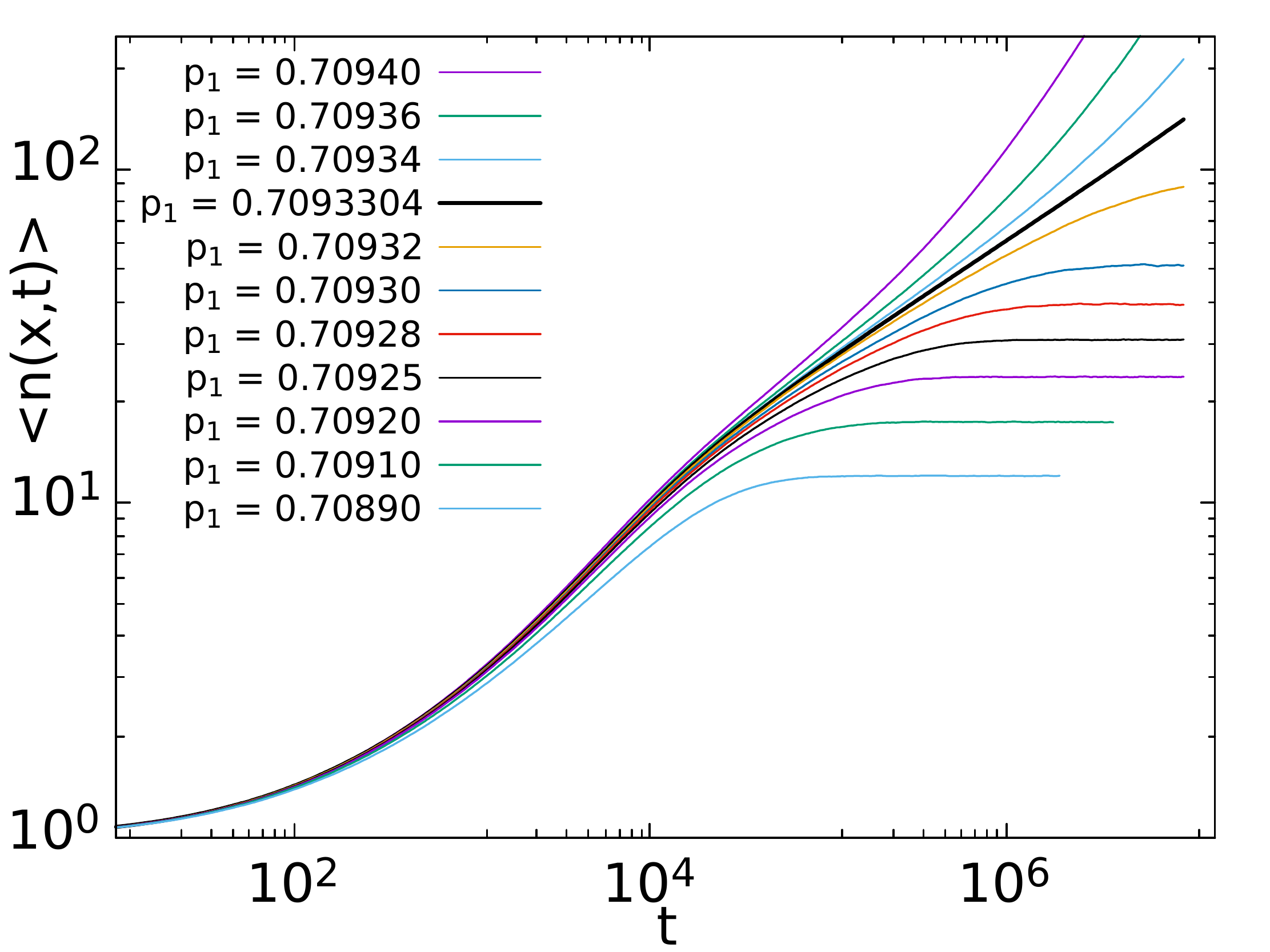}
\par\end{centering}
        \caption{\label{fig:n_t_B}  Average heights for $p_1 \approx p_{1,c}$ at point
        B (easy detachment) in the presence of a barrier, plotted against $t$.}
\end{figure}

\begin{figure}
\begin{centering}
        \includegraphics[scale=0.35]{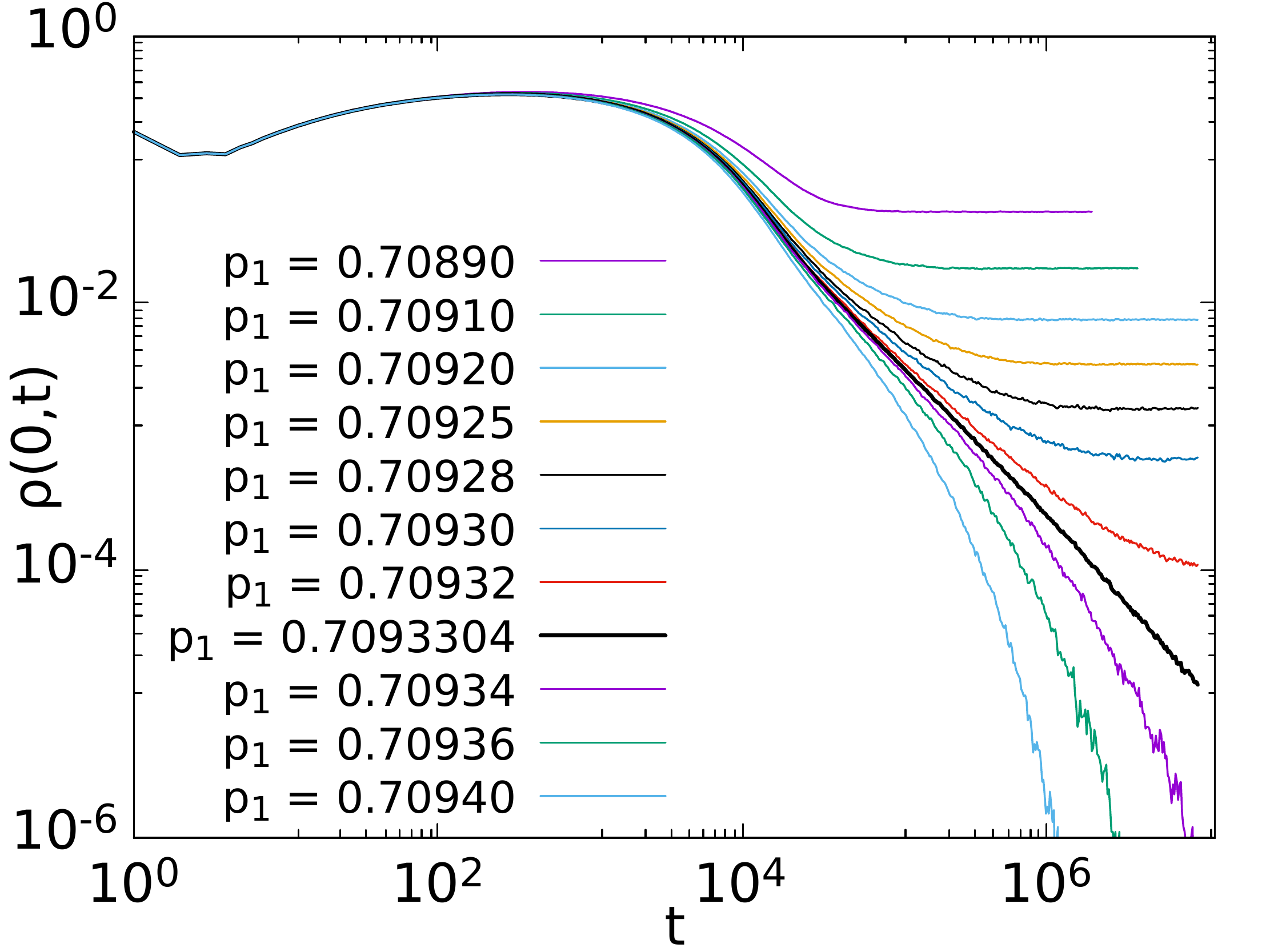}
\par\end{centering}
        \caption{\label{fig:P_0-B}  Fractions of the interface that are at height $n=0$ for
        several values of $p_1$, at point B and in the presence of a barrier.}
\end{figure}

\begin{figure}
\begin{centering}
        \includegraphics[scale=0.35]{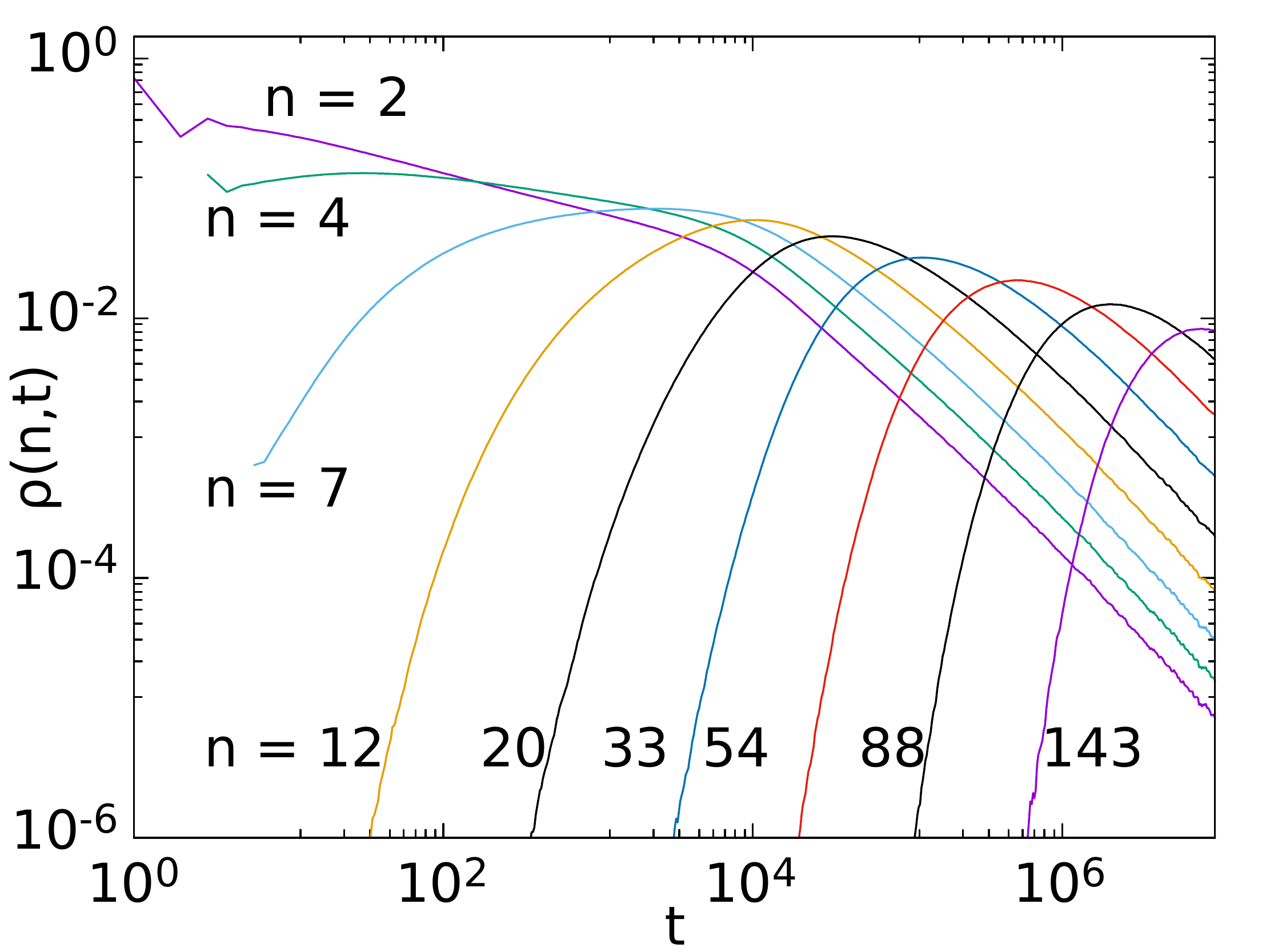}
\par\end{centering}
        \caption{\label{fig:B_crit-rho_nt}  Fractions of the interface that are at heights 
	$n=2,\;4,\;7,\;12 \ldots 143$ for $p_1=p_{1,c}$, at point B and in presence of a barrier.}
\end{figure}

For point B, a plot similar to Fig.~\ref{fig:n_A} gives $\alpha = 0.47$ and $p_{1,c} = 0.7093304(2)$.
At this point, we made exactly the same simulations and plots that we had also made at point A. The results
are shown in Figs.~\ref{fig:n_t_B}, \ref{fig:P_0-B} and \ref{fig:B_crit-rho_nt}. 

We see from all three figures that corrections to scaling are now even bigger than for uneasy detachment.
But, fortunately, in spite of them we can definitely exclude that easy and uneasy detachment are in the 
same universality class. The clearest indication is from the average heights shown in 
Figs.~\ref{fig:n_t_A} and \ref{fig:n_t_B}. While the former shows for the critical curve a consistent 
downward curvature, giving thus an upper bound $\mu \leq 0.23(1)$ for uneasy detachment, the critical
curve in Fig.~\ref{fig:n_t_B} shows for large $t$ an {\it upward} curvature, giving thus $\mu \geq 0.38(1)$ 
for easy detachment.

The same conclusion is reached by comparing Figs.~\ref{fig:P_0-A} and  \ref{fig:P_0-B}. For large $t$,
the critical curve in the former is bent {\it upward}, leading to $\delta \leq 0.60(3)$, and a careful
inspection shows that it is bent {\it downward} in Fig.~\ref{fig:P_0-B}, giving rise to $\delta \geq 1.24(3)$
for easy detachment.

\section{Relation to avalanche propagation in the directed Oslo model \label{sec:DOM}}

 In sandpile type models, the  critical density in the steady state  is decided by the requirement that   the average height behind a propagating  avalanche
is same as the average height in front of it. 
How the avalanches develop when the density is exactly critical depends on details of correlations in the critical background.  In the following, we study the propagation of avalanches on specially prepared backgrounds and try to identify when  the background  becomes critical,  and  then study the universality classes of  this spreading process in  a specific case: the directed Oslo rice-pile   model in two dimensions \cite{Oslo-Dhar}.  
 We will show that for a class of specially prepared backgrounds, the avalanche propagation in a version of the directed Oslo ricepile model becomes equivalent to the interface model studied above.
The central object of study in the rice pile model \cite{Oslo} is also an interface and its height $h$,
but this is {\it not} identified with the interface $n(x,t)$.
Here we have a two dimensional  square lattice of size $2L \times T$ with periodic boundary conditions in $x$ and
open boundary condition in $t$. In a stable configuration, each site $(x,t)$ can contain up to two grains of rice.
More precisely, call $h(x,t)$ the number of grains at site $(x,t)$, and define at each site a
critical height $h_c(x,t)$. It is initially set for each $(x,t)$ to an independent random value
with probabilities ${\rm prob}[h_c=2]=p$ and ${\rm prob}[h_c=3]=1-p$. A configuration is stable iff
$h(x,t) < h_c(x,t)$ at each site.

Let us start with a random stable configuration, where each site is attributed 
independently a height 0, 1, or 2 with probabilities $c_0, c_1, $ and $c_2$, with $c_0+c_1+c_2=1$. 
Then we add one grain of rice at a randomly chosen
$x$ and at $t=0$. If this leads to an unstable configuration (i.e., if this site then has
$h(x,t) \geq h_c(x,t)$), the site {\it topples}. During a toppling, one grain falls from $(x,t)$ to
$(x-1,t+1)$, and another from $(x,t)$ to $(x-1,t+1)$. This may lead to instablities at $(x-1,t+1)$ 
or $(x-1,t+1)$, which lead then to further topplings and to the evolution of an avalanche. In the 
usual mode of operation of the model, critical heights at toppling sites are re-set randomly 
(again with probabilities $p$ and $1-p$), and when an avalanche is finished, a new avalanche starts.
Here we modify also this aspect of the model: We randomly re-set all critical and actual heights,
before a new avalanche starts. Thus each avalanche evolves in a new {\it uncorrelated} background,  
with critical height probabilities controlled by $p$ and actual height probabilities controlled by
$c_i$.

\begin{figure}
\begin{centering}
        \includegraphics[scale=0.5]{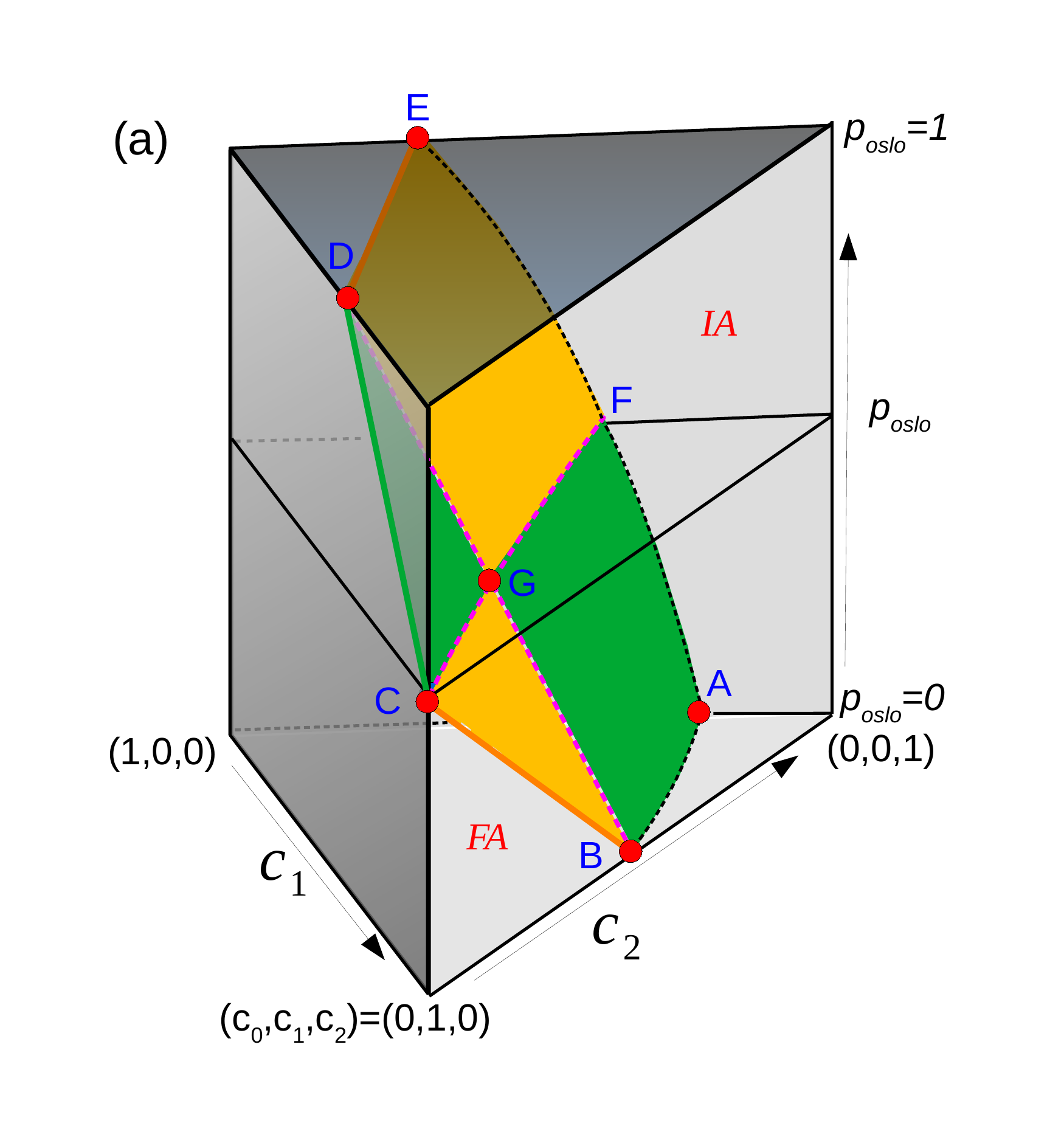}
        \includegraphics[scale=0.35]{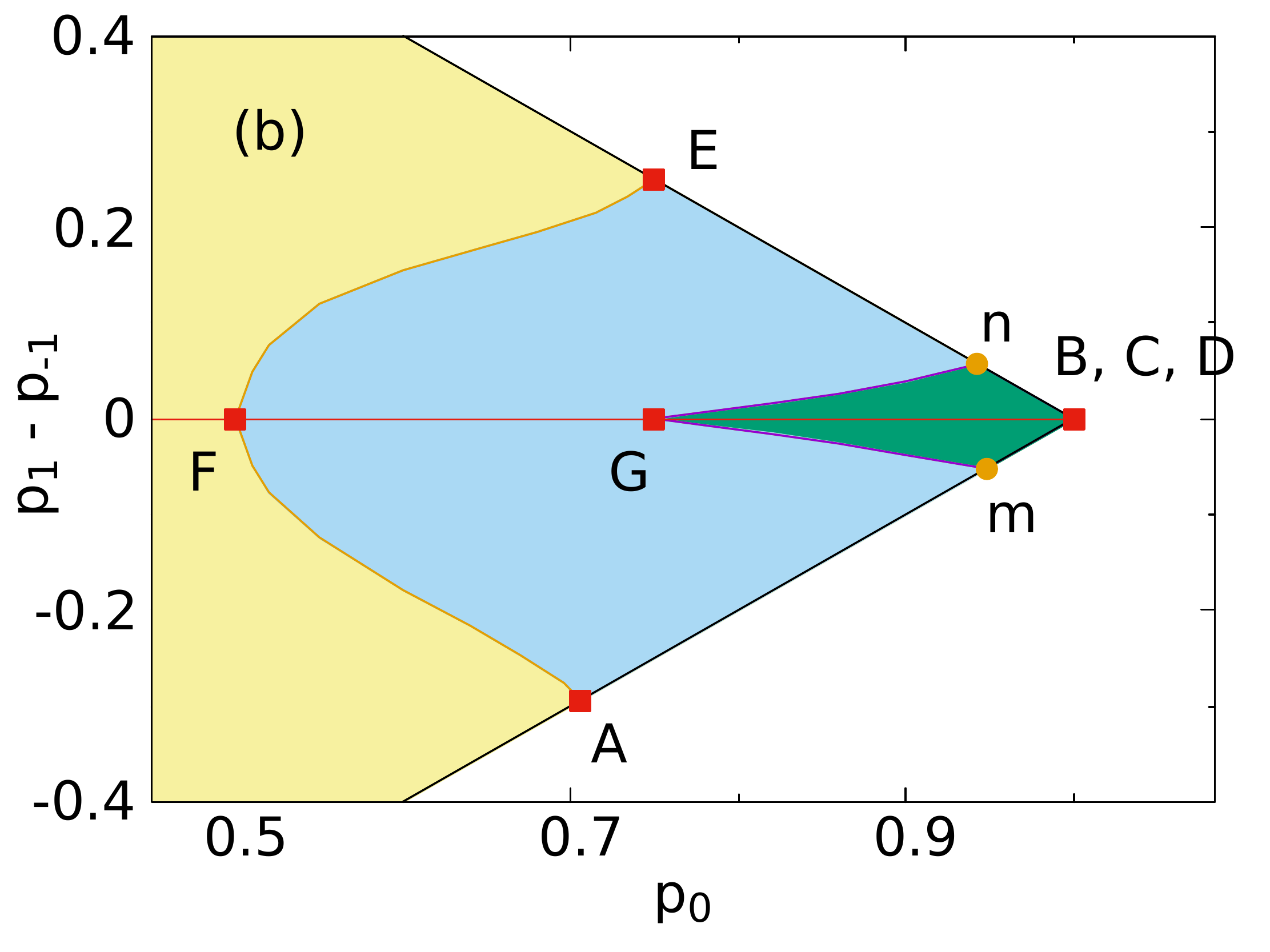}
\par\end{centering}
	\caption{\label{Oslo}  (a) Rough sketch of the phase portrait of the 
	directed Oslo model. Finite avalanches ("FA") are left of or  below the surface spanned by the bounding curve
	  ABCDEFA, while infinite avalanches ("IA") can occur to its right or above. The line CGF 
	is precisely at $p=p_{oslo}=1/2$. On it the model obeys detailed balance, and it is 
	exactly the original directed Oslo model with the original mode of operation, while 
	avalanches evolve always in uncorrelated backgrounds off this line. In the green regions 
	of the critical surface (including the boundary lines AB and CD), the transition is DP, 
	while it is the tent transition in the two yellow regions. More precisely, along the lines 
	AB and CD the model maps onto the DK model, and point A corresponds precisely to directed
	site percolation.
	(b) Region in the interface model parameter space onto which the critical surface of 
	panel (a) is mapped. The green region is covered three times by this map, while the blue 
	area is covered once, and the yellow area is not covered at all.}
\end{figure}

We identify $n(x,t)$ in the interface model with the number of topplings in the Oslo sandpile  at the 
site $(x,t)$. Then $n_{in}(x,t)$ is the number of grains that come to $(x,t)$.  Clearly, $n(x,t)$ is 
zero, if $n_{in}(x,t)$ is zero. Now consider the case when $n_{in}(x,t)$ is odd. Since the site would 
have $0,1 $ or $2$ grains, with probabilities $c_0, c_1, c_2$, we see that the number of topplings 
would be $(n_{in}(x,t)+1)/2$, with probability $ [c_1 p+c_2]$, and otherwise the number is 
$ [ n_{in}(x,t)-1]/2$. Similarly, if $n_{in}(x,t) $ is a positive even number, the number of topplings 
at $(x,t)$ is $n_{in}/2-1, n_{in}/2, n_{in}/2 +1$ with probabilities 
$c_0 ( 1 -p), c_0 p + c_1 + c_2 ( 1-p), c_2 p$ respectively. This gives us the identification
\bea \label{map}
  &&p_0=p c_0+ c_1 + (1-p)c_2, \nonumber \\
  && q= p c_1 +c_2,  \\
  && p_1=pc_2.  \nonumber
\eea
Notice that the mapping Eq.~(\ref{map}) from $(c_i,p)$ to $(p_i,q)$ is neither invertible nor
onto. To see the latter, notice that the regions $q< p_1$ and $p_0+p_1 < q$ do not correspond to any
$(c_i,p)$ with $0 \leq c_i\leq 1, 0 \leq p\leq 1,$ and $\sum_i c_i=1$. A more careful analysis shows
that this is also true for the whole region $p_0 < 1/2-(p_1-p_{-1})^2/2$. For the lack of invertibility,
define first
\be
    r = p_0 + 2p_1 + q.      \label{def_r}
\ee
Using Eq.~(\ref{map}) one finds that $r$ can also be written as $r=c_1+2c_2+p$, and that
\bea
    q &=& p(r-2c_2-p) + c_2 \nonumber \\
      &=& pr-p^2 + (1-2p)\times(p_1/p) .
\eea
Inserting this into Eq.~(\ref{def_r}) we obtain then a cubic equation for $p$,
\be
    p^2(r-p)-p(r-p_0) +p_1 = 0.   \label{cubic}
\ee
Depending on the values of $p_0,p_1$, and $q$ this can have one, two, or three real-valued solutions,
from which $c_1$ and $c_2$ can be computed by $c_2=p_1/p$ and $c_1=r-2c_2-p$, except when $p=0$. In
the latter case we have from Eq.~(\ref{map}) $c_2=q$ and $c_1=p_0-c_2$.

The boundary between finite and infinite avalanches in the $(c_i,p)$ space is sketched in 
Fig.~\ref{Oslo}(a). It is the surface bounded by the  bounding curve  ABCDEFA. Avalanches are always finite in 
the lower left of it, while infinite avalanches can occur in the upper right. The region in the $(p_i,q)$
space onto which the surface spanned by  ABCDEFA is mapped is shown in Fig.~\ref{Oslo}(b) (green and blue 
areas). The green area is covered three times by the map (i.e., each point is the image of three points 
in Fig.~\ref{Oslo}(a)), while the blue area is covered once. Notice that the lines BG, CG, and DG 
in Fig.~\ref{Oslo}a are all mapped onto the same line $p_1=p_{-1}=0$ in  Fig.~\ref{Oslo}(b). 

In addition, we have the following comments:

First,  along the line AB we have $p=0$ and therefore $p_1=0$, according to Eq.~(\ref{map}). The latter is 
also true along the line CD, since there $c_2=0$. Thus both AB and CD map onto the DK model. Furthermore, 
in point A we have also $c_1=0$ and therefore $q=p_0$. As we had seen in Sec. IV.B3, this maps onto 
directed site percolation.

Second,  similarly, one sees that both lines CB and DE correspond to $p_1$ =0, and thus to the tent transition.

Third, by continuity, it follows that the entire green regions in Fig.~\ref{Oslo}(a) correspond to DP, while 
the transition is of tent type in the yellow regions.

Fourth, the lines mG and nG in Fig.~\ref{Oslo}(b) have two preimages each. One preimage of point `m' is 
on the line AB, the other is on CD. Similarly, the point `n' has one preimage on DE, the other on BC.

Fifth, let us finally discuss the special case with $c_1=1/2, c_0=p/2, c_2=(1-p)/2$, for any $p\in (0,1)$.
This is precisely the line BGD in Fig.~\ref{Oslo}(a). On the one hand, it maps into the same line 
$p_1=p_{-1}$ in Fig.~\ref{Oslo}(b) as the line CGF. On the other hand,
it can be proven that the steady state of the directed rice pile model in the usual mode of operation
(i.e., without the random re-setting after each avalanche) is a
product state with precisely these probabilities. From Eq.~(\ref{map}) we obtain then
\be
    q=1/2,\; p_1=p_{-1} = p(1-p)/2.
\ee
With these parameters, 
the background  after the avalanche has gone through is statistically equivalent to the initially prepared  background, and thus, this line  correspond to the SOC state. 
Thus we see that the interface model with these parameters is mapped {\it exactly} onto the unmodified directed rice pile model \cite{Oslo-Dhar}.

\section{Discussion and Conclusions}

We have introduced and studied -- mostly by numerical simulations -- a simple 1D non-equilibrium
model that involves three control parameters. As happens in many similar 1D models, it allows for more 
than one physical interpretation. Indeed, the number of different interpretations is bewildering large,
and the number of different critical phenomena which it displays in various regions of control parameter
space is fascinating.

The most natural interpretation is as a model for growing interfaces, similar to the famous KPZ model, but 
with a lower barrier. When the interface hits the barrier it gets stuck to it, and it can detach from it only 
at the borders (`fronts') of the attached regions. In this sense, it is similar to the models in
\cite{Pikovsky,Munoz,Shimla,Hinrichsen-Livi}, but the details are very different. In particular, the 
activity in \cite{Pikovsky,Munoz,Shimla,Hinrichsen-Livi} is in general real-valued, while it is discrete
in our model. More specifically, the models in \cite{Pikovsky,Munoz,Shimla,Hinrichsen-Livi} can also be 
formulated as models with multiplicative noise, which is impossible for integer-valued activity.

Another large class of phenomena to which our model bears some resemblance is wetting and de-wetting
\cite{Starov,de_Gennes2}. But again, the analogies are far from perfect -- mostly because the special 
discrete nature of our model implies discrete contact angles. Indeed, depending on control 
parameters, the `liquid' wetting the base (barrier) can have either a flat or a triangle-shaped surface.
Nevertheless, our model shows a feature that resembles the distinction between high and low wettability:
Lateral spreading of the activity below interfaces that are partially attached to 
the barrier can be `pulled' or `pushed' \cite{Saarloos}. 

This distinction establishes a relation to another large class of phenomena: Front propagation into unstable 
states \cite{Saarloos}. But it is also of fundamental importance in our model, since we can interpret the 
model -- in the case of pulled fronts, and only then -- as a realization of the important contact process 
(directed percolation) universality class \cite{Hinrichsen}. Thus we have also an immediate relationship 
with epidemic spreading. But, compared to all previous realizations of epidemic spreading based on the 
contact process where sites can  be only active or inactive, our present realization allows for varying and 
unbounded local activity. Epidemic spreading with non-trivial 
local activity levels is one of the basic features of helminth infections \cite{helminth}. It would be
interesting if our model could indeed be applied to them.

While the lateral spreading of attached interfaces is thus related to and relevant for epidemic spreading,
the detachment of interfaces from the barrier is most interesting for wetting and multiplicative noise
models. In our model, non-attached interfaces are in general of KPZ type, implying that they are 
typically not symmetric under up or down reflection. More precisely, 1D interfaces of KPZ type can be 
either noisy concatenations of cup-convex arcs or of cap-convex arcs. We found that this distinction
has definitely consequences for the detachment transition: In both cases we found critical behavior 
(i.e., the transition is continuous), but the two cases are in different universality classes. We 
reached this conclusion (which is opposite to claims in \cite{Munoz,Hinrichsen-Livi}, which were 
however made for a similar class of models)
in spite of very large corrections to scaling, which made it impossible to give precise estimates for 
critical exponents. But we could give rather precise bounds, such that the upper bounds for one case
were lower than the lower bounds for the other.

We should, finally, point out that large scaling corrections are, maybe, the most pervading feature 
of the present study, and are responsible for most uncertainties that remain. In part, they are not
unexpected: Since robust universality classes like the directed percolation class are attractive 
in the renormalization group (RG) sense, we must expect large cross-over phenomena due to RG flows from 
other (unstable) fixed points. But this is not all. In addition we found that the expected KPZ scaling for 
free (non-attached) interfaces sets in very late, due to a special feature of our model \cite{Grass-KPZ}, 
namely that the interface propagation velocity depends periodically on the interface tilt.

There are a number of open questions. One is the precise scaling for the detachment transitions.
The other is the precise way that KPZ scaling is reached, in particular when the control parameter $q$
is close to 1/2, which is also the region where the KPZ scaling should cross over to Edwars-Wilkinson
scaling. Due to the latter difficulty, e.g., the scaling for what we called the `clipped KPZ' transition
is not yet well understood. But similar uncertainties remain also in other regions of the control 
parameter space. For instance, the curves in Fig.~1 seem to approach the lines $p_1=0$ and $p_1=1$
tangentially, and merge with them at non-trivial points. Within numerical uncertainty, we cannot
however exclude the possibility that they meet these lines only at the end point $p_0=1$. Another
open problem is the precise relationship with the directed Oslo rice pile model sketched in Sec.  VI. This relationship is well understood at the symmetry line $q=1/2.$  The other regions of parameter space do not have any direct relationship to the steady state behavior of the ricepile model.

In this paper, we have dealt with one very specified simplified model 
which showed the entire bouquet of fascinating features. But how robust are they?
Otherwise said: Are there similar models, either with continuous variables or on other lattices, which 
display the same features? In particular, do models exist which exhibit most of these features, but without a periodic tilt dependence of interface velocities?

Finally, the last and maybe most difficult open question is whether   generalizations exist to higher dimensions with similarly rich behavior.

\vglue .5cm 
  \begin{acknowledgments}
  P.G. thanks the Leverhulme Trust for financial support, and the University of Aberdeen --
  where part of this work was done -- for a most pleasant stay. D.D.'s research was supported by a Senior Scientist fellowship by the National Academy of Sciences of India.
  
  \end{acknowledgments}


\begin{thebibliography}{99}

\bibitem{Reichl} L. Reichl, {\it A Modern Course in Statistical Physics}, 3rd ed. (Wiley-VCH, New York, 2016).

\bibitem{Castellano} C. Castellano, S. Fortunato, and V. Loreto, ``Statistical physics of social dynamics", 
   Rev. Mod. Phys. {\bf 81}, 591 (2009).

   \bibitem{Bouchaud} J. -P. Bouchaud and M. Potters, {\it Theory of financial risk and derivative pricing:
   from statistical physics to risk management}, 2nd edition (Cambridge Univ. Press, Cambridge 2003).

   \bibitem{deGennes} P. G. de Gennes, {\it Scaling concepts in polymer physics} (Cornell Univ. Press, Ithaca 1979).

   
   \bibitem{epidemics} N. T. J. Bailey, {\it The mathematical theory of infectious diseases and its applications}.
	2nd edition (Griffin, London 1975)
	
 \bibitem{Stauffer} D. Stauffer and A. Aharony {\it Introduction to Percolation Theory}, 2dn Edition 
   (Taylor \& Francis, London 1994).

   \bibitem{Grass-Sund} P. Grassberger and K. Sundermeyer, ``Reggeon field theory and Markov processes", 
   Phys. Lett. B {\bf 77}, 220 (1978).

   \bibitem{Sugar} J. L. Cardy and R. L. Sugar, ``Directed percolation and Reggeon field theory", J. Phys. A: 
   Math. Gen. {\bf 13} (1980) L423 (1980).

   \bibitem{Paczuski} M. Paczuski and S. Boettcher, ``Universality in Sandpiles, Interface Depinning, and Earthquake 
   Models", Phys. Rev. Lett. {\bf 77}, 111 (1996).

   \bibitem{btw} P. Bak, C. Tang and K. Wiesenfeld, ``Self-organized criticality: An explanation of the 1/f noise",
   Phys. Rev. Lett. {\bf 59} (1987) 381; ``Self-organized criticality'', Phys. Rev. {\bf A 38} (1988) 364.

   \bibitem{Oslo-Dhar} D. Dhar and R. Ramaswamy, ``Exactly solved model of self-organized critical phenomena",
   Phys. Rev. Lett. {\bf 63}, 1659 (1998).

   \bibitem{Oslo} V. Frette, K. Christensen, A. Malthe-S{\o}rensen, J. Feder, T. J{\o}ssang, and P. Meakin,
   ``Avalanche dynamics in a pile of rice", Nature {\bf 379}, 49 (1996).

   

	\bibitem{helminth} R. M. Anderson, ``Population dynamics and control of hookworm and roundworm infections", in
	{\it Population Dynamics of Infectious Diseases: Theory and Applications} (ed. R.M.Anderson), pp. 67-108
        (Chapman \& Hall, London 1982).

        \bibitem{helminth1} Ascaris\_lumbricoides. (n.d.) In Wikipedia. Retrieved December 15, 2022, from 
	https://en.wikipedia.org/wiki/Ascaris\_lumbricoides

	\bibitem{helminth2} L. S. Roberts and J. Janovsky (Jr.), {\it Foundations of Parasitology}, 8th edition 	(McGraw-Hill, 2009).

	\bibitem{KPZ} M. Kardar, G. Parisi, and Y. -C. Zhang, ``Dynamic scaling of growing interfaces", 
   Phys. Rev. Lett. {\bf 56}, 889 (1986).

   \bibitem{EW} S. F. Edwards and D.R. Wilkinson, ``The surface statistics of a granular aggregate", Proc. R. Soc. 
   London, Ser. {A381}, 17 (1982).

   \bibitem{Barabasi} A. -L. Barabasi and H. E. Stanley, {\it Fractal Concepts in Surface Growth}, pp. 123-125 
	(Cambridge University Press, Cambridge, UK, 1995).

	\bibitem{Grass-KPZ} P. Grassberger, ``Kardar-Parisi-Zhang type dynamics with periodic tilt dependence of the 
	propagation velocity in 1+1 dimensions", arXiv:2111.11414 (2021)

	\bibitem{grassberger_percol_high-d} P. Grassberger, ``Critical percolation in high dimensions", Phys. Rev. E {\bf 67},
        036101 (2003).

        \bibitem{grassberger_log4d} P. Grassberger, ``Logarithmic corrections in (4+1)-dimensional directed percolation",
        Phys. Rev. E {\bf 79}, 052104 (2009).

        \bibitem{Foster} J. G. Foster, P. Grassberger, and M. Paczuski, ``Reinforced walks in two and three dimensions",
        New Journal of Physics {\bf 11}, 023009, (2009).

        \bibitem{JfWang} J. Wang, Z. Zhou, Q. Liu, T. M. Garoni, and Y. Deng, ``High-precision Monte Carlo study of directed
        percolation in (d+1) dimensions", Phys. Rev. E {\bf 88}, 042102, (2013).

       
        
        \bibitem{Devillard} P. Devillard and H. Spohn, `` Universality class of interface growth with reflection symmetry",
        J. Stat. Phys. {\bf 66}, 1089 (1992).
\bibitem{Saarloos} W. van Saarloos, ``Front propagation into unstable states", Phys. Rep. {\bf 386}, 29 (2003).

\bibitem{Starov} V. M. Starov, M. G. Velarde, and C. J. Radke, {\it Wetting and Spreading Dynamics} (CRC Press, Taylor \& Francis Group, New York, 2007). 

\bibitem{Torcini} A. Torcini, P. Grassberger, and A. Politi, ``Error propagation in extended chaotic systems",
    J. Phys A: Math. Gen. {\bf 28}, 4533 (1995).

    \bibitem{Essam} J. W. Essam, ``Directed compact percolation: cluster size and hyperscaling",
	J. Phys. A {\bf 22}, 4927 (1989).

	\bibitem{domany} E. Domany and W. Kinzel, ``Equivalence of cellular automata to Ising models and directed percolation",
       Phys. Rev. Lett. {\bf 53}  311 (1984).
  
       \bibitem{improvedH1} M. Campostrini, M. Hasenbusch, A. Pelissetto, P. Rossi, E. Vicari, "Critical exponents and equation of state of the three-dimensional Heisenberg universality class", Phys. Rev. B {\bf 65}, 144520 (2002)
   
       \bibitem{improvedH2} N. Clisby, "High resolution Monte Carlo study of the Domb-Joyce model", J. Phys.: Conf. Ser. {\bf 921}, 012012 (2017).
       
     
       \bibitem{Jensen} I. Jensen, ``Low-density series expansions for directed percolation: I. A new efficient 
	algorithm with applications to the square lattice", J. Phys. A: Math. Gen. {\bf 32}, 5233 (1999).

	\bibitem{Kashchiev} D. Kashchiev, {\it Nucleation} (Butterworth-Heinemann, Oxford 2000).

	\bibitem{ball} J. M. Ball, J. Carr, and O. Penrose, ``The Becker-Doering cluster equations: Basic properties and 
    asymptotic behaviour of solutions", Comun. Math. Phys. {\bf 104}, 657 (1986).

    \bibitem{cahn} J. W. Cahn, ``Critical point wetting", J. Chem. Phys. {\bf 66}, 3667 (1977).

    \bibitem{coinfect} P. Grassberger, L. Chen, F. Ghanbarnejad, and W. Cai, ``Phase transitions in cooperative 
   coinfections: Simulation results for networks and lattices", Phys. Rev. E {\bf 93}, 042316 (2016).

   \bibitem{footnote2} Notice that we would have expected {\it a priori} an exponent $\alpha = 1/3$; see Ref.\cite{Grass-KPZ}.

   \bibitem{Pikovsky} A. Pikovsky and J. Kurths, ``Roughening interfaces in the dynamics of perturbations of 
   spatiotemporal chaos", Phys. Rev. {\bf E49}, 898 (1994).

   \bibitem{Munoz} M. A. Mu\~noz, G. Grinstein, and Yuhai Tu, ``Survival probability and field theory in systems 
   with absorbing states", Phys. Rev. E {\bf 56}, 5101 (1997).

   \bibitem{Shimla} P. Grassberger, ``Directed Percolation: Results and Open Problems", in {\it Nonlinearities in 
   complex systems}, Proc. 1995 Shimla Conf. on Complex Systems (Narosa Publishing, New Dehli, 1997).

   \bibitem{Hinrichsen-Livi} H. Hinrichsen, R. Livi, D. Mukamel, and A. Politi, ``Model for Nonequilibrium Wetting 
   Transitions in Two Dimensions", Phys. Rev. Lett. {\bf 79}, 2710 (1997).

   \bibitem{de_Gennes2} P. -G. de Gennes, F. Brochard-Wyart, and D. Qu\'er\'e, {\it Capillarity and Wetting Phenomena} 
	(Springer, New York 2004).

	\bibitem{Hinrichsen} H. Hinrichsen, ``Non-equilibrium critical phenomena and phase transitions into absorbing 
	states", Advances in Physics {\bf 49}, 815 (2000).
\end{thebibliography}
\end{document}